\documentclass[11pt]{article}
\usepackage{hyperref}
\usepackage{amsmath, amsfonts, amssymb, mathrsfs}
\usepackage{dcolumn}
\usepackage{caption}
\usepackage{subcaption}
\usepackage{filemod}
\usepackage[ruled, vlined]{algorithm2e}
\usepackage{floatrow}
\usepackage{setspace}
\usepackage{verbatim}
\usepackage[Export]{adjustbox}
\usepackage{tabularx}
\usepackage{cellspace}
\usepackage{mathtools}
\usepackage{natbib}
\usepackage[title]{appendix}
\usepackage{bbm}
\usepackage{multirow}
\usepackage[dvipsnames]{xcolor}
\hypersetup{
    colorlinks=true,
    linkcolor=blue, 
    urlcolor=black,
    citecolor=blue, 
    }

\title{\vspace{-0.3cm} Modernizing full posterior inference 
for surrogate modeling of categorical-output simulation experiments}
\author{Andrew Cooper\thanks{Corresponding author: Department of Statistics, 
	Virginia Tech, {\tt ahcooper@vt.edu}} 
	\and Annie S. Booth\thanks{Department of Statistics, Virginia Tech} 
	\and Robert B. Gramacy\footnotemark[2]}
\date{\today}

\begin{document}

\vspace{-0.5cm}
\maketitle

\singlespacing

\vspace{-1.5cm}
\begin{abstract} 
Gaussian processes (GPs) are powerful tools for nonlinear classification in
which latent GPs are combined with link functions.  But GPs do not scale well
to large training data.  This is compounded for classification where the
latent GPs require Markov chain Monte Carlo integration. Consequently, fully
Bayesian, sampling-based approaches had been largely abandoned. Instead,
maximization-based alternatives, such as Laplace/variational inference (VI)
combined with low rank approximations, are preferred. Though feasible for
large training data sets, such schemes sacrifice uncertainty quantification
and modeling fidelity, two aspects that are important to our work on surrogate
modeling of computer simulation experiments.  Here we are motivated by a large
scale simulation of binary black hole (BBH) formation. We propose an
alternative GP classification framework which uses elliptical slice sampling
for Bayesian posterior integration and Vecchia approximation for computational
thrift.  We demonstrate superiority over VI-based alternatives for BBH
simulations and other benchmark classification problems.  We then extend our
setup to warped inputs for ``deep'' nonstationary classification.
\end{abstract}


\singlespacing

\section{Introduction}
\label{sec:1}

Model-based simulations of real-world phenomena allow researchers to conduct
experiments that are either too costly or infeasible to perform in field tests
\citep{Morris1993,santner2018design}; however, even simulations can be too
computationally intensive to repeatedly sample.  Meta-modeling a
simulator with a statistical replacement -- or ``surrogate'' -- trained on a
collection of simulator runs (i.e., input-output pairs) 
enables estimation of simulator output for unobserved inputs at a
significantly reduced computational cost \citep{gramacy2020surrogates}. An
effective surrogate is able to capture the important features of a
simulator's response surface, while providing appropriate uncertainty
quantification (UQ) at locations where training data is lacking. This can be
crucial to many otherwise simulation-intensive downstream tasks such as
computer model calibration \citep{kennedy2001bayesian}, input sensitivity
analysis \citep{marrel2009calculations}, and active learning or
Bayesian optimization \citep{jones1998efficient}.

One popular surrogate is a Gaussian process \citep[GP; see,
e.g.,][]{Rasmussen2006}, which models the response as multivariate normal
(MVN) with a covariance based on pairwise distances between inputs. GPs can
flexibly capture nonlinear dynamics within a Bayesian framework,
thereby providing predictive mean and variance with intuitive and well-calibrated UQ. Training a
GP and evaluating its predictive equations involves decomposing covariance
matrices, requiring flops that are cubic in the number of observations, $n$.
This $\mathcal{O}(n^3)$ computational bottleneck can limit GP surrogates
to small-to-medium sized simulation campaigns, e.g., $n \ll 10{,}000$.

Our work is motivated by a large simulation campaign studying binary black
holes \citep[BBHs;][]{Lin2021, yazdi2024deep} that can form when two massive
celestial objects collide.  The simulator estimates the ``chirp mass'' of the
resulting BBH depending on eleven inputs describing the configuration of the
system and celestial objects. BBH formation is rare; for most input
configurations the simulator returns {\tt NA}, meaning no BBH was formed.
Here, we study the important sub-problem of predicting whether or not a BBH
forms for a given input configuration.  
We propose tackling this problem with a GP surrogate, which presents two,
intertwined challenges.  The first is adapting GP surrogates to the
classification problem while preserving modeling fidelity and UQ. The second
is handling large data sizes,  i.e., ``big $n$.'' 

The typical setup for GP-based classification (GPC) analogues the way logistic
regression extends linear regression, where an $n$-dimensional latent layer
is estimated through an appropriate link function to a Bernoulli response
(more on this in Section \ref{sec:class_review}).  This setup is intuitive but
necessitates inference for at least $n$-many unknown quantities.
Fully Bayesian GPC via Markov chain Monte
Carlo (MCMC) posterior integration for all unknowns
had long been abandoned, unless $n \ll 1{,}000$. Instead,
practitioners preferred thriftier alternatives such as approximate
variational inference \citep[VI;][]{damianou2016},
typically in tandem with inducing points \citep[IPs; e.g.,][]{quinonero2005},
for a low-rank matrix representation, and thus speedier matrix decompositions.
But this recipe can lead to underwhelming predictive performance
because the shortcuts are too severe \citep{wu2022variational}.
\citet{Sauer2022} argue that in the context of computer simulation experiments
(with real-valued outputs), VI with IPs also undercuts UQ.  
We see similarly poor behavior in our surrogate classification setting,
which we will demonstrate empirically later.

To extend the toolkit for fully Bayesian GPC, we follow the roadmap provided
by \citet{Sauer2022}, for ``deep'' GP regression, and adapt it to classification.
This involves two ingredients: elliptical slice sampling
\citep[ESS;][]{murray2010} for high-dimensional posterior integration of
Gaussian latent quantities and Vecchia approximation for sparse decomposition
\citep{vecchia1988estimation, Katzfuss2021,datta2016hierarchical}.  ESS for
Bayesian GPC is not new; in fact, that was the algorithm's motivating
application.  Nevertheless, this approach is too cumbersome without additional
approximation. Our novel insight is that Vecchia approximation is uniquely
suited to ESS for GPC.   We show that our Vecchia/ESS GPC provides more
accurate predictions, with better UQ, compared to VI/IP and similar
maximization-based alternatives, and provides similar performance to
non-Vecchia-approximated ESS when possible, in smaller-$n$ settings.  We then
take things one step further and show how the deep, warping-layer appoach of
\citeauthor{Sauer2022}, for regression, can be ported to classification for
improved performance in nonstationary settings.

The paper is laid out as follows. Section \ref{sec:back} reviews GP modeling.
Section \ref{sec:method} begins by extending that review to include historical
context for state-of-the-art large scale GPC, ultimately arguing that Vecchia
with ESS has the potential to provide high-fidelity, fully Bayesian inference
without compromise. Section \ref{sec:implement} provides implementation
details, illustrations on pedagogical examples, and ultimately application on
the BBH simulator. Section \ref{sec:deepgp} introduces deep GP extensions and
additional benchmarking. Section \ref{sec:discuss} concludes with a discussion
and directions for future work.

\section{Background}
\label{sec:back}

Here we provide a foundational description of GP surrogate modeling, first in
the context of regression before extending to classification. We also
provide a survey of modern GPC methods in order to frame our main
contributions beginning in Section \ref{sec:method}.

\subsection{Gaussian process regression}
\label{sec:gp_review}

Consider a black-box simulator $y = f(x)$, where $x\in\mathbb{R}^d$ is the
input, and $y\in\mathbb{R}$ is the output. Let $X = [x_1^\top,\dots,x_n^\top]$
be an $n \times d$ matrix collecting $n$ inputs, and let $Y
=(y_1,\dots,y_n)^\top$ be a column $n$-vector of outputs. A GP model assumes
$Y\sim\mathcal{N}_n\left(0,\Sigma(X)\right)$, where $\Sigma(X)$ is an $n\times
n$ covariance matrix built using rows of $X$. The construction of $\Sigma(X)$
depends on our choice of \textit{kernel}, $\Sigma(x_i, x_j) \equiv
\Sigma(X)^{ij}
\equiv \Sigma(X, X)^{ij}$, which determines the pairwise correlation between
outputs $y_i$ and $y_j$, often in terms of (inverse) distance between their
inputs. One common choice is the \textit{squared-exponential} kernel,
\begin{equation}
\Sigma(x, x') = \tau^2\exp\left\{-\dfrac{||x - x'||^2}{\theta}\right\}. \label{eq:sigma}
\end{equation}
Our work here is largely agnostic to kernel choice.  Many kernels 
\citep{abrahamsen1997review} have forms similar to Eq.~(\ref{eq:sigma}), where
hyperparameters $\theta$ and $\tau^2$ determine the  
distance--correlation strength (lengthscale) and the overall 
amplitude of the function (scale), respectively.

The MVN form of $Y \mid X$ elicits a likelihood, say for any hyperparameters
in $\Sigma$, that is proportional to
\begin{equation}
\mathcal{L}(Y \mid X) \propto |\Sigma(X)|^{-\frac{1}{2}}\exp\left\{-\frac{1}{2}Y^\top\Sigma(X)^{-1}Y\right\}.
\label{eq:like}
\end{equation}
One may maximize (the $\log$ of) Eq.~(\ref{eq:like}) with respect to $\theta$
and $\tau^2$ to obtain point estimates \citep{gramacy2020surrogates}. We instead opt
for a Bayesian approach that combines information from the likelihood with
that from a prior. For $\tau^2$, an independent inverse-Gamma prior is
conjugate, and in fact can be marginalized out analytically
\citep{gramacy2020surrogates}. Unfortunately there is no known conjugate prior
choice for $\theta$; Metropolis-Hastings (MH) can be used to obtain
samples from its posterior \citep{hastings1970}.

Settings for these hyperparameters are rarely of direct interest; rather
their value lies in how they capture
input--output dynamics through $\Sigma(X)$ and how that extends to {\em
predictions} for novel inputs $\mathcal{X} \in\mathbb{R}^{n'\times d}$.  First
consider fixed hyperparameter values, either MLE estimates or a
single sample from the posterior.  Let
$\Sigma(\mathcal{X})$ be defined analogously to $\Sigma(X)$ via pairs in
$\mathcal{X}$, both tacitly conditioned on the same hyperparameters.  
Similarly populate
$\Sigma(\mathcal{X}, X)$, an $n' \times n$
matrix with $(ij)^{th}$ entry defined by the kernel calculation between the $i^\mathrm{th}$ row of
$\mathcal{X}$ and $j^\mathrm{th}$ row of $X$.  A joint model between observed
$Y$ and unknown $\mathcal{Y}$ is $(n + n')$-variate MVN with mean zero and
covariance matrix composed block-wise of $\Sigma(X)$, $\Sigma(\mathcal{X})$,
$\Sigma(\mathcal{X}, X)$, and $\Sigma(\mathcal{X}, X)^\top$.  Standard MVN
conditioning yields
\begin{equation}
\begin{split}
\mathcal{Y}(\mathcal{X}) \mid X,Y \sim \mathcal{N}_{n'}\left(\mu_\mathcal{Y}(\mathcal{X}), 
\Sigma_\mathcal{Y}(\mathcal{X})\right)\text{,\hspace{1mm} where \hspace{2mm}} &
\mu_\mathcal{Y}(\mathcal{X}) = \Sigma(\mathcal{X},X)\Sigma(X)^{-1}Y\\
&\Sigma_\mathcal{Y}(\mathcal{X}) = \Sigma(\mathcal{X})-
\Sigma(\mathcal{X},X)\Sigma(X)^{-1}\Sigma(\mathcal{X},X)^\top.
\end{split}
\label{eq:pred}
\end{equation}
These are sometimes referred to as the {\em kriging equations}
\citep{matheron1963}.  In a Bayesian setting, we may filter posterior samples
$\{(\theta^{(t)}, \tau^{2(t)})\}_{t=1}^T$ through Eq.~(\ref{eq:pred}) to
obtain samples for predictive moments $\mu_{\mathcal{Y}}^{(t)}(\mathcal{X})$
and $\Sigma_{\mathcal{Y}}^{(t)}(\mathcal{X})$.  These may in turn be used to
generate $\mathcal{Y}^{(t)}$ samples via MVN draws, providing a Monte Carlo
(MC) approximation to the posterior for $\mathcal{Y}$. Each batch of
hyperparameters requires new calculation of covariance matrices, with
subsequent inversion of $\Sigma(X)$ at $\mathcal{O}(n^3)$ computational cost.

\subsection{Gaussian process classification}
\label{sec:class_review}

Now suppose outputs are binary, i.e., $y_i \in \{0, 1\}$ for $i =
1,\dots,n$, with some probability $p(x_i)$ of $y_i = 1$, and $1-p(x_i)$
of $y_i = 0$.  Our earlier development for $f(x): \mathbb{R}^d \rightarrow \mathbb{R}$
is not readily applicable as a model for probabilities $p(x) \in [0,1]$.
But we can follow the generalized linear modeling
\citep[GLM; e.g.,][]{McCullagh:1989} approach of deploying an {\em inverse-link function}, 
$p(x_i) = \sigma(f(x_i))$ for any monotonic sigmoid $\sigma: \mathbb{R}\rightarrow [0,1]$.  
In our work we privilege the canonical logit link, whose inverse is the logistic function 
$\sigma(z) = \dfrac{1}{1 + e^{-z}}$; however, our development is
not tied to this particular choice.  
With $y_i \sim \mathrm{Bern}(\sigma(z_i))$,
posterior inference for {\em latent} $Z$-values may commence
through the Bernoulli likelihood
\begin{equation}
\mathcal{L}(Y \mid Z) = \prod_{i=1}^n \sigma(z_i)^{y_i}\left(1-\sigma(z_i)\right)^{1-y_i}.
\label{eq:class_like}
\end{equation}
Bayesian formulations hinge on the prior for $Z$.
\citet{albert1993} chose linear $Z\mid X \sim \mathcal{N}_n(X\beta, \mathbb{I}_n\tau^2)$,
where data augmentation and Gibbs samplers work well \citep{fruhwirth2010data}.

With a GP prior, e.g., $Z \mid X \sim \mathcal{N}_n(0,\Sigma(X))$, 
Bayesian inference is similar but fraught with
challenges because everything involving $Z$ requires cubic-in-$n$ flops. 
Elliptical slice sampling \citep{murray2010} was
developed to extend the Gibbs-like posterior sampling framework from Bayesian
linear classification to GPs.  We shall return to ESS in Section
\ref{sec:ess}, for now remarking that fully Bayesian posterior sampling of an
$n$-variate latent $Z$ is considered cumbersome because it exacerbates cubic
computational issues.  

Early practitioners of GP classification preferred point-inference techniques
based on the Laplace approximation \citep{williams1998} and the like.  This
approach subsequently became enshrined as canonical for GP classification in
machine learning \citep[][Chapter 3]{Rasmussen2006}. Things have evolved
somewhat over the last twenty years; the standard benchmark is now a
variational approximation \citep[e.g.,][]{damianou2016} with a low rank,
inducing points-based GP covariance
\citep{Snelson2006,titsias2009,banerjee2013}. More recently, a ``doubly
stochastic'' variational inference (DSVI) method has become popular for both
ordinary and ``deep'' GPC modeling \citep{salimbeni2017}. The downside to
point-wise inference is reduced UQ, as happens when integration is
replaced with maximization.  When UQ is essential, such as in active learning
for computer experiments, authors have gone to great lengths to avoid {\em
both} point-inference and MCMC \citep{grapol2011} while remaining
computationally tractable.  Approximating the covariance structure helps with
computation, but low-rank approximations still blur predictions and sacrifice
fidelity \citep{wu2022variational}.

\subsection{An illustration}
\label{sec:class_ex}

\begin{figure}[ht!]
\centering
\includegraphics[scale=0.65,trim=0 10 0 40,clip=TRUE]{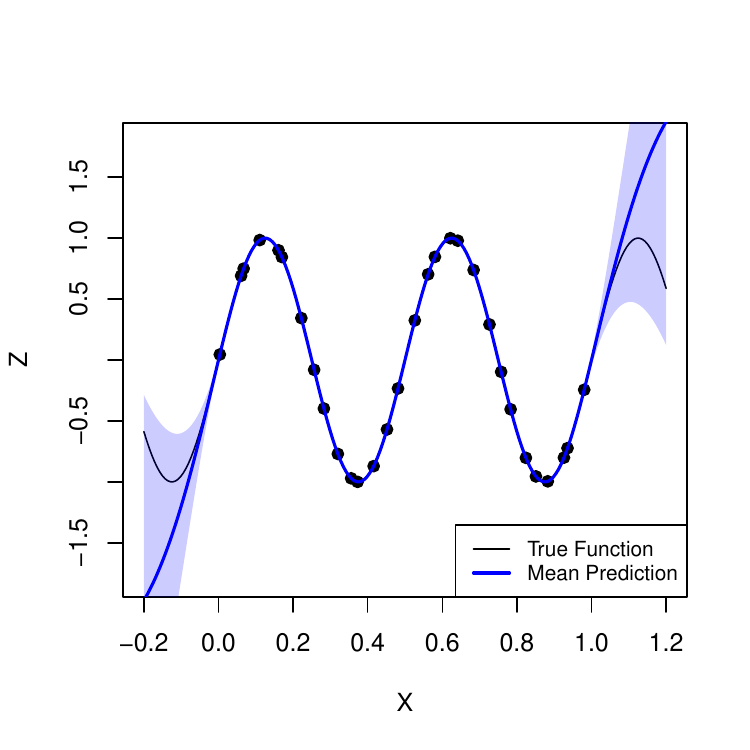}
\includegraphics[scale=0.65,trim=0 10 0 40,clip=TRUE]{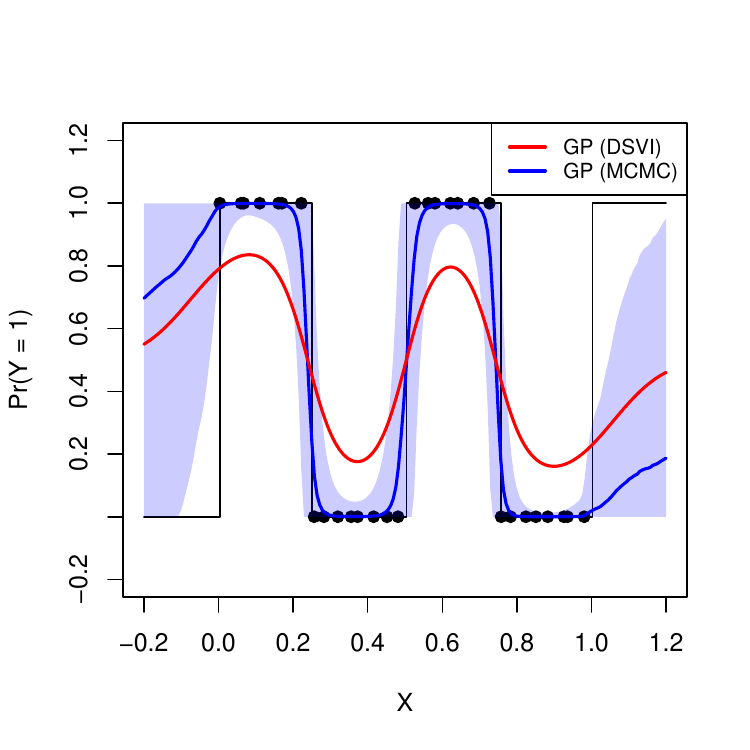}
\caption{GP predictions from regression and classification models. Real-valued
response data $Y$ from a sine function (\textit{left}) is ``binarized'' into
0's and 1's based on the boolean $Y > 0$ (\textit{right}).  Fully Bayesian
GP(C) shown in blue with 95\% prediction interval shaded; DSVI GPC shown in red.}
\label{fig:1d_ex}
\end{figure}

In order to illustrate several of these concepts at once, Figure
\ref{fig:1d_ex} provides a caricature contrasting GP regression (left) with
GP logistic classification (right), where the latter uses a latent $Z$ following
the same sinusoid as the former. 
In the classification context the $Z$-values remain unobserved. Consequently,
there can be a great degree of posterior uncertainty in these quantities, both
in- and out-of-sample, as indicated by the wider prediction intervals in the
right panel compared to the left, both from fully Bayesian fits.  The
nonparametric nature of the GP prior on $Z$ allows for more flexibility than a
linear model, which would struggle to go ``back-and-forth'' between the
alternating regimes without a basis expansion.  This makes GP inference
conceptually more nuanced than the (Bayesian) linear case, even if
operationally they are similar because they involve an MVN prior. For
classification, the fully Bayesian GP (blue) gives ``crisper'' probabilities
that the DSVI GP (red).  Note that the library we used for DSVI, described
later, does not furnish a predictive interval for the latent quantities; i.e.,
no UQ.

\section{Modernizing an old idea for Bayesian GPC}
\label{sec:method}

We wish to make predictions
for an unknown label $y$ at a novel input $x$, through
probabilities $p_y(x) \equiv \mathbb{P}\left(y(x) = 1\right)$, while
marginalizing over all {\em a posteriori} possible latent $z$-values:
\begin{align}
p_y(x) = \int_{\mathbb{R}} \sigma(z(x)) \cdot p(z(x) \mid X, Y )\; d z &= 
\int_{\mathbb{R}} \int_{\mathbb{R}^n} \sigma(z(x)) \cdot p(z(x), Z \mid X, Y )\; dZ dz \nonumber \\
&= \int_{\mathbb{R}} \int_{\mathbb{R}^n} \sigma(z(x)) \cdot p(z(x) \mid Z, X, Y) \cdot p(Z \mid X, Y) \; dZ dz. \label{eq:gpcpost}
\end{align}
The integral above is written to express the simplest case, univariate
 over a single $z$ at arbitrary $x$.  Our focus is on the
inner-integral calculating $p(z(x) \mid X, Y )$ by marginalizing over the
posterior for $Z$ given training data $(X,Y)$. Conditional on $Z$, $p(z(x)
\mid Z, X)$ is a straightforward application of GP prediction via
Eq.~(\ref{eq:pred}), albeit with $Z$ instead of $Y$, which can be performed
separately for each $x$ of interest, or jointly for $\mathcal{Y}$ via
$\mathcal{Z}$ over a set of $\mathcal{X}$'s.  Hyperparameters, such as
lengthscales $\theta$, would also require posterior marginalization. We
first focus on integrating over $p(Z \mid X, Y)$ in Sections
\ref{sec:hist}--\ref{sec:vecchia}, and then return to prediction and hyperparameter
inference in Section \ref{sec:latent}.

Marginalizing over $p(Z\mid X, Y)$ requires integrating over $\mathbb{R}^n$, and there is no known
closed form.  VI/Laplace methods work around this via approximation combined
with optimization. We reviewed some of these in Section \ref{sec:class_review}
while \citet[][Section 3.3, pp.,~41]{Rasmussen2006} provide a more complete
list. The reason for VI's popularity is complicated, and has as much to do
with circumstance as with technology at the turn of the $21^{\mathrm{st}}$
century.  So we indulge in a quick historical digression to help frame our
novel contribution in spite of many elements being established over a decade
ago.

\subsection{Historical digression}
\label{sec:hist}

The state-of-the-art in fully Bayesian GPC at the end of the
$20^{\mathrm{th}}$ century involved Metropolis samplers
\citep{bernardo1998regression}.  Although tuning and blocking helped, these
generally produced sticky chains and slow convergence in $n$-dimensional
latent $Z$ space and were only practical for $n < 100$ or so.  Laplace methods
\citep{williams1998} and those based on expectation propagation
\citep{minka2001family}, a variant of VI, broke through that barrier in a big
way.  About a decade later \citet{murray2010} solved the MCMC/Metropolis
mixing problem for latent GPs, of which GPC was one example. Despite the
simplicity of their ESS algorithm [Section \ref{sec:ess}] it didn't catch on.
MCMC via ESS was slower than Laplace/VI, and as problems got bigger ($n \gg
1{,}000$) cubic-in-$n$ complexity meant that thousands of MCMC iterations
-- even with excellent mixing -- couldn't compete with mere dozens of
   Newton-like iterations.  It was better to get something in a reasonable
   amount of time than to get nothing after waiting forever.

As problems became even bigger ($n \gg 10{,}000$) many
practitioners leveraged sparse or low-rank approximations to $\Sigma$
\citep[e.g.,][]{melkumyan2009sparse,emery2009kriging,gramacy2015local,cole2021locally,quinonero2005unifying,lazaro2010sparse}
to avoid cubic bottlenecks. One particular low-rank approximation, now known as inducing points, turned out to be particularly well-suited for VI.
But IP approximations offer low-resolution in large $n$ settings; where a
wealth of data ought to reveal a rich and nuanced response surface, GPC via VI/IP
instead oversmooths, like in Figure \ref{fig:1d_ex}.

Recently, there has been renewed interest in the \citet{vecchia1988estimation}
approximation for GP regression as an alternative to IPs
\citep[e.g.,][]{katzfuss2020,datta2021sparse,stein2004,stroud2017bayesian,datta2016hierarchical,Katzfuss2021,katzfuss2022scaled}.
It turns out, as we will show shortly in Section \ref{sec:vecchia}, Vecchia
is ideally suited to (by now long forgotten) ESS for GPC. We are not the first
to apply Vecchia in the context of GPC, or other latent Gaussian models
\citep[e.g.,][]{zilber2021,Cao2023}, but we are the first to do so in a fully
Bayesian context via ESS.

\subsection{Elliptical slice sampling}
\label{sec:ess}

The hard part of Eq.~(\ref{eq:gpcpost}) is integrating $p(Z \mid X, Y)$ over
$Z \in \mathbb{R}^n$, which is high dimensional and
analytically intractable.  Bayes rule provides $p(Z \mid X,Y) \propto p(Y \mid
Z) \cdot p(Z \mid X)$ where $p(Y \mid Z)$ is the Bernoulli likelihood 
from Eq.~(\ref{eq:class_like}), and $p(Z \mid X)$ is the
GP prior $Z\sim\mathcal{N}_n(0,\Sigma(X))$, which is not conjugate for that
likelihood. 
\citeauthor{murray2010}'s elliptical slice sampling
algorithm is designed exactly for this setting: MVN prior with
arbitrary likelihood.  ESS is inspired by (ordinary) slice sampling
\citep[SS;][]{neal2003slice} which is more general in one sense (any
prior/target distribution) but more specific in another (one-dimensional
sampling).

Algorithm \ref{alg:ess} provides the details.  It is worth clarifying that any
log likelihood $\ell(\cdot)$ may be used, with any MVN prior $\mathcal{N}(\mu,
\Sigma)$. ESS is not particular to GPC;  \citet{murray2010} included GPC as an
example among others. Like SS, ESS is a ``rejection-free'' method in that only
a single random proposal $Z' \sim \mathcal{N}_n(0,\Sigma(X))$ is required, and
a quantity calculated from $Z'$, and possibly other random scalars, is
eventually returned.  It establishes a Markov chain because the returned
quantity $Z^{(t+1)}$ is a function of both $Z'$ and the previous sample
$Z^{(t)}$. In contrast to Metropolis samplers, $Z^{(t+1)}$ is never a copy of
the previous sample $\left(Z^{(t+1)} \ne Z^{(t)}\right)$. Like rejection
sampling \citep{robert1999}, ESS has a loop over random deviates, but unlike
rejection sampling each iteration increases the probability of ultimate
acceptance.

\medskip
\begin{algorithm}[ht!] 		
\DontPrintSemicolon
{\bf Input:} Previous $Z^{(t)}$, response $Y$, covariance $\Sigma(X)$,  log likelihood  $\ell$. \\ 
{\bf Output:} Posterior sample $Z^{(t+1)} \sim p(Z|X,Y)$. \\
~

Draw $Z'\sim\mathcal{N}_n(0,\Sigma(X))$. \;
Draw $u\sim\text{Unif}[0,1]$ and set acceptance threshold $\ell_{\text{thresh}} = \ell(Y \mid Z^{(t)}) + \log(u)$. \;
Draw angle $\gamma\sim\text{Unif}[0,2\pi]$. Construct bracket $\gamma_{\text{min}}=\gamma-2\pi$, $\gamma_{\text{max}}=\gamma$. \;
~

\While{1}{
Calculate proposal $Z^* = Z^{(t)}\cos(\gamma) + Z'\sin(\gamma)$ and evaluate $\ell_{\text{prop}} = \ell(Y \mid Z^*)$. \;
\If{$\ell_{\text{prop}} > \ell_{\text{thresh}}$}{Return $Z^{(t+1)} = Z^*$.}
\Else{
\If{$\gamma < 0$}{$\gamma_{\text{min}}=\gamma$}
\Else{$\gamma_{\text{max}}=\gamma$}
Draw new $\gamma\sim\text{Unif}[\gamma_{\text{min}},\gamma_{\text{max}}]$.
}
}
\caption{Elliptical slice sampling for GPC estimation.}
\label{alg:ess}
\end{algorithm}

\begin{figure}[ht!]
\centering
\includegraphics[scale=0.55,trim=0 10 0 50,clip=TRUE]{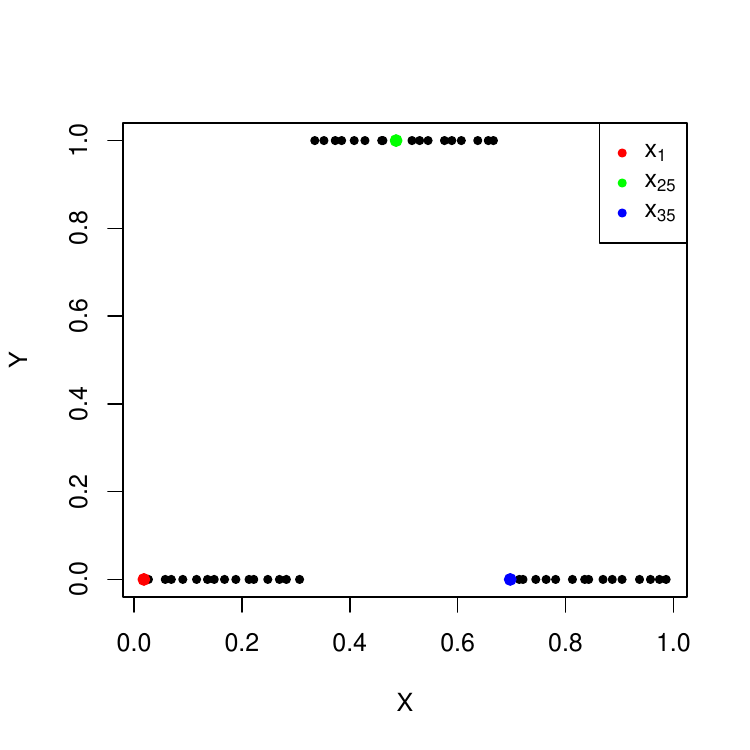}
\includegraphics[scale=0.55,trim=0 10 0 50,clip=TRUE]{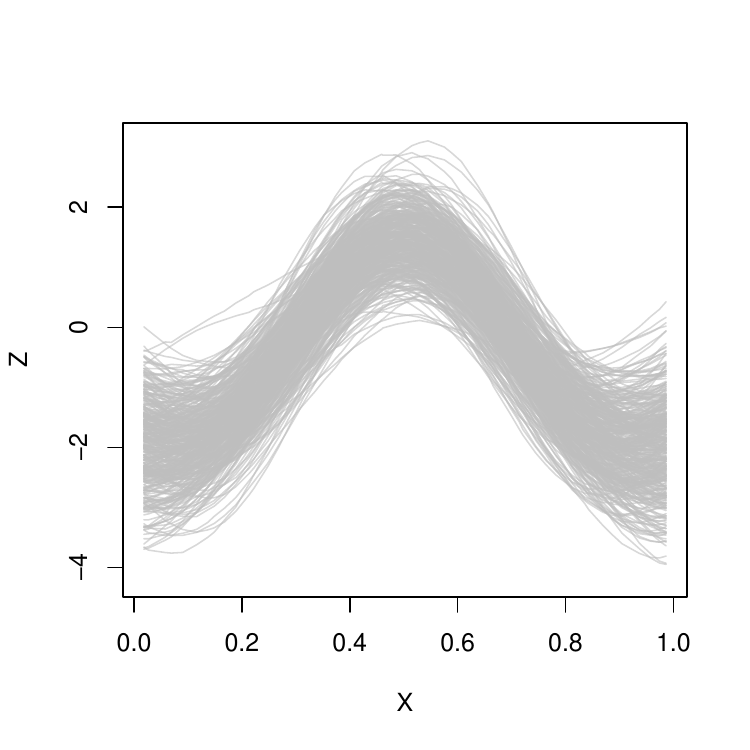}
\includegraphics[scale=0.55,trim=0 10 0 50,clip=TRUE]{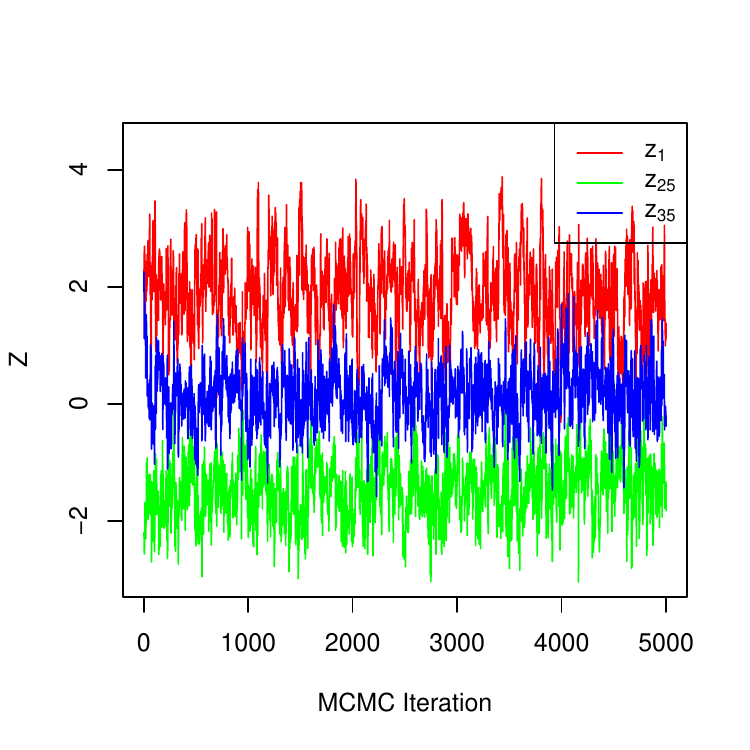}
\includegraphics[scale=0.55,trim=0 10 0 50,clip=TRUE]{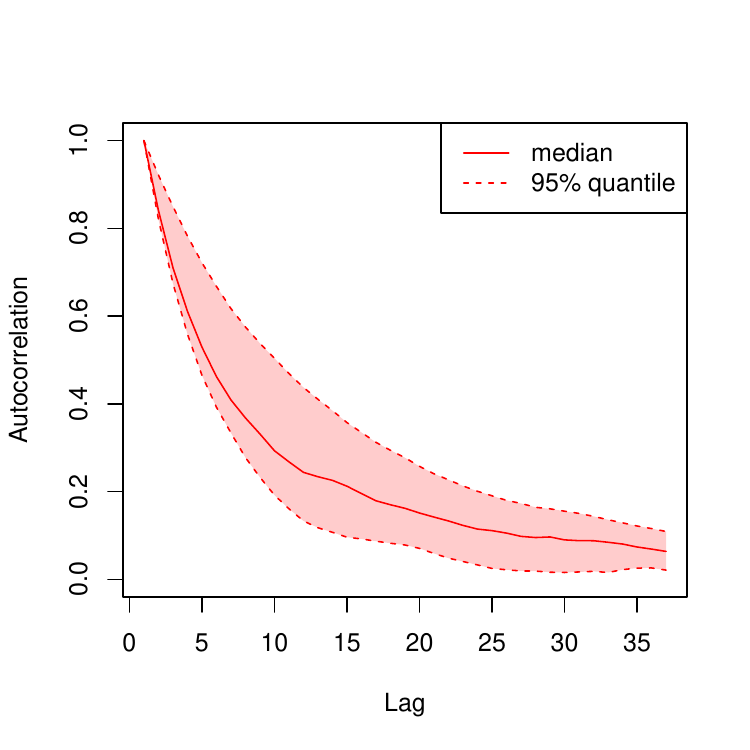}
\caption{ESS GPC illustration. {\em Top-left:} training data with 
``top-hat'' response. {\em Top-right:} samples of $Z^{(1:T)}$ after burn-in
and thinning.  {\em Bottom-left}: chain of $z^{(t)}$ values at $x_1$, 
$x_{25}$ and $x_{35}$. {\em Bottom-right:}
autocorrelation  distribution (median and 95\%
quantiles) for all $Z^{(1:T)}$.}
\label{fig:ess}
\end{figure}
To illustrate ESS for GPC, consider the ``top-hat'' example
\citep{dunlop2018deep} shown in the top-left panel of Figure \ref{fig:ess}.
The response steps from $y(x_i)=0$ to $1$ and back again as $x_i$ varies from
left to right. We observe $Y$ values at a Latin hypercube sample
\citep[LHS;][]{mckay1979comparison} $X$ of size $n=50$. Our GP prior assumes
$Z\sim\mathcal{N}(0,\Sigma(X))$, where $\Sigma(X)$ follows the
squared-exponential kernel (\ref{eq:sigma}) with fixed $\theta = 0.1$. We
obtain $T=5{,}000$ posterior samples from $Z \mid X,Y$ following
Alg.~\ref{alg:ess} for $t=1,\dots,T$, beginning with $z_i^{(1)} =
\log\left(-2\tau\right)$ if $y_i = 0$, and $z_i^{(1)} =
\log\left(2\tau\right)$ if $y_i = 1$, where $\tau$ is chosen via a heuristic
discussed later in Section \ref{sec:imp_details}. This took about five seconds
on a conventional, single-core workstation, or about one thousandth of a
second per draw. Each ESS sample required around six loop iterations on
average, but sometimes as few as one or as many as fourteen. The remaining
samples after a burn-in of $1{,}000$ and a thinning rate of $10$ are shown as
gray lines in the top-right panel of the figure. Notice each $Z^{(t)}$ is a
smooth, continuous realization.  The bottom-left panel shows raw chains,
samples $z_i^{(1)},\dots,z_i^{(T)}$, for three indices $i\in
\{1,25,35\}$.  These showcase the GPC's varying confidence in
a 1-label depending on an input's closeness to transitions between classes.
Lastly, the bottom-right panel provides the median and $95\%$ interval for the
autocorrelation of our samples across all $n=50$ training observations. These
showcase the sampler's excellent mixing and ability to avoid getting
``stuck,'' as a Metropolis sampler might.

Besides its simplicity, perhaps the most important takeaway from
Alg.~\ref{alg:ess} -- which we repeat here for emphasis -- is that only one
MVN instance $Z'$ is required for each $Z^{(t)}
\rightarrow Z^{(t+1)}$.  Besides evaluating log likelihoods, which for our
Bernoulli GPC setup requires flops linear in $n$, the only real work lies in
generating $Z' \sim \mathcal{N}_n(0, \Sigma(X))$.  If you were looking to make 
ESS for GPC more computationally efficient, say, to handle larger $n$ in a 
reasonable amount of time, this is where you would target your effort. 
In Section \ref{sec:bbh} we explore the BBH example with $n$ in the tens of 
thousands, which is not tractable for even one MCMC iteration, let alone 
hundreds or thousands, no matter how good ESS mixing is.  

Although there are many algorithms for MVN sampling, the most common is via
Cholesky decomposition. \citet{gelman1995} explain that one may convert $n$
independent standard normal deviates $\xi = (\xi_1, \dots \xi_n)$ into MVN
ones via $Z' = \mu + U
\xi$, where $U$ is a Cholesky factor for $\Sigma$, or $\Sigma(X) = UU^\top$.
Calculating $U$ is cubic in $n$ for dense matrices $\Sigma$.  The Vecchia
approximation \citep{vecchia1988estimation}, discussed next, may be the most
expedient way of generating a sparse Cholesky factor for covariance matrices
$\Sigma(X)$ based on inverse distance.   Although Vecchia-based GP
approximation has seen a resurgence of interest lately for GP
likelihoods (\ref{eq:like}) and predictive equations (\ref{eq:pred}), as we
have already reviewed, we are the first to recognize its value in the context
of GPC via ESS, where all that's required are MVN proposals.

\subsection{Vecchia approximation}
\label{sec:vecchia}

At its core, \citet{vecchia1988estimation}'s idea for approximating MVN
likelihoods is simple.  Any joint probability distribution may be formulated
as a product of cascading conditionals $p(z_1, z_2, z_3) = p(z_3 \mid z_2,
z_1) p(z_2 \mid z_1) p(z_1)$, in any order or indexing.  
An approximation may be obtained by dropping some of the conditioning 
variables, e.g., $p(z_1, z_2, z_3) \approx p(z_3 \mid z_1)
p(z_2 \mid z_1) p(z_1)$.  This is applicable to any distribution, but the 
quality of the approximation depends on many factors.  In the case of MVNs, 
two remarks are noteworthy: (i) those conditional distributions are (univariate) 
Gaussian and have a convenient closed form (\ref{eq:pred});
(ii) the computations within this closed form will involve smaller matrices
if conditioning variables are reduced. 

Although that description is intuitive, modern Vecchia for GPs is not done
through products of conditional distributions.  Rather, computations
leverage the fact that dropping conditioning variables induces conditional
independence. In other words, $p(z_3 \mid z_1) p(z_2 \mid z_1) p(z_1)$ encodes
that $z_3$ is independent of $z_2$ given $z_1$.  Conditional independence
equates to zeros in the inverse covariance (a.k.a., precision) matrix.
A modern Vecchia approximation induces sparsity in the MVN precision
matrix for an $n$-dimensional $Z$. 

Most recent literature on Vecchia
\citep[e.g.,][]{katzfuss2020,datta2016hierarchical} focuses on
aspects of the approximation that target the regression GP setting, including
its effect on the the full, joint likelihood (\ref{eq:like}), and other
aspects of Bayesian inference including predictive equations (\ref{eq:pred}).
These details do not pertain directly to the classification context, at least
when ESS is used, although they shall be relevant later when we discuss
broader aspects of our sampling procedure in Section \ref{sec:latent} and our
deep GP enhancements in Section \ref{sec:deepgp}.  As we explain in Section
\ref{sec:ess}, the most expensive operation in ESS sampling
for GPC is generating MVN proposals.  Here we leverage a result from
\citet{Katzfuss2021}, which shows how the precision matrix can be written as an
upper--lower Cholesky decomposition: $\Sigma(X)^{-1} = U U^\top$.

Let $c(i) \subseteq \{1,\dots, i-1\}$ denote the conditioning set for the
$i^{\mathrm{th}}$ variate in the MVN for $i=1,\dots, n$, i.e., describing the
univariate conditional $z_i \mid Z_{c(i)}$.  Also let $X_{c(i)}$ denote the rows of
$X$ corresponding to $Z_{c(i)}$ for a GP.  \citeauthor{Katzfuss2021} show how
entries of the $n \times n$ triangular matrix $U$ can be calculated as
\begin{equation}
 U_X^{ji} = \begin{cases} 
      \dfrac{1}{\sigma_i(X_{c(i)})} & i = j \\
      -\dfrac{1}{\sigma_i(X_{c(i)})}\Sigma\left(x_i, X_{c(i)}\right)\Sigma\left(X_{c(i)}\right)^{-1}[\text{index of } j \in c(i)] & j\in c(i) \\
      0 & \text{otherwise,}
   \end{cases}
\label{eq:u_mat}
\end{equation}
where $\sigma_i^2(X) \equiv \Sigma_{\mathcal{Z}}(x_i) = 
\Sigma(x_i)-\Sigma\left(x_i,X\right)\Sigma\left(X\right)^{-1}
\Sigma\left(X,x_i\right)$, i.e., following Eq.~(\ref{eq:pred}) with
singleton $\{x\} = \mathcal{X}$. Several
details are worth noting here.  First, $U^{ji}$ entries may be populated 
in parallel as they only depend on $X$, conditioning indices $c(i)$, and 
covariance hyperparameters. 
In our implementation (more details in Section \ref{sec:imp_details}) we 
populate this matrix via {\tt OpenMP} for symmetric multi-core speedups.  
If we limit the conditioning set size to a maximum of $m$ variables 
($|c(i)| \leq m$), then each $U^{j\cdot}$ row can be calculated in 
time $\mathcal{O}(m^3)$, owing primarily to
the cost of decomposing the $(m
\times m)$-dimensional matrix $\Sigma(X_{c(i)})$.  Populating all $n$ rows of
$U$ requires flops in $\mathcal{O}(nm^3)$, which is a substantial improvement
over the usual $\mathcal{O}(n^3)$ cost of a Cholesky decomposition 
when $m\ll n$.

For our purposes (ESS for GPC), this means the following:  proposals $Z'
\sim \mathcal{N}_n (0, \Sigma(X))$ may be drawn by first sampling $a_1, \dots,
a_n \stackrel{\mathrm{iid}}{\sim} \mathcal{N}(0, 1)$, then calculating $Z' =
(U^\top)^{-1}a$. A simple rearrangement yields the form $U^TZ' = a$, where
$Z'$ can be found via a sparse forward solve and, crucially, without the need
to directly invert $U$ \citep{Sauer2022}.  In subsequent ESS iterations, if
hyperparameters are fixed -- or if they are otherwise identical to those used
in the previous draw because a Metropolis proposal for the lengthscale was
rejected -- one may proceed without re-calculating $U$.  The quality of the
approximation depends on the conditioning set size, $m$, and the selection of
$c(i)$.  We follow \citeauthor{Sauer2022}~in choosing $m=25$ and additional
specifications provided along with other implementation details in Section
\ref{sec:implement}.

\begin{figure}[ht!]
\centering
\includegraphics[scale = 0.50,trim=15 10 20 25]{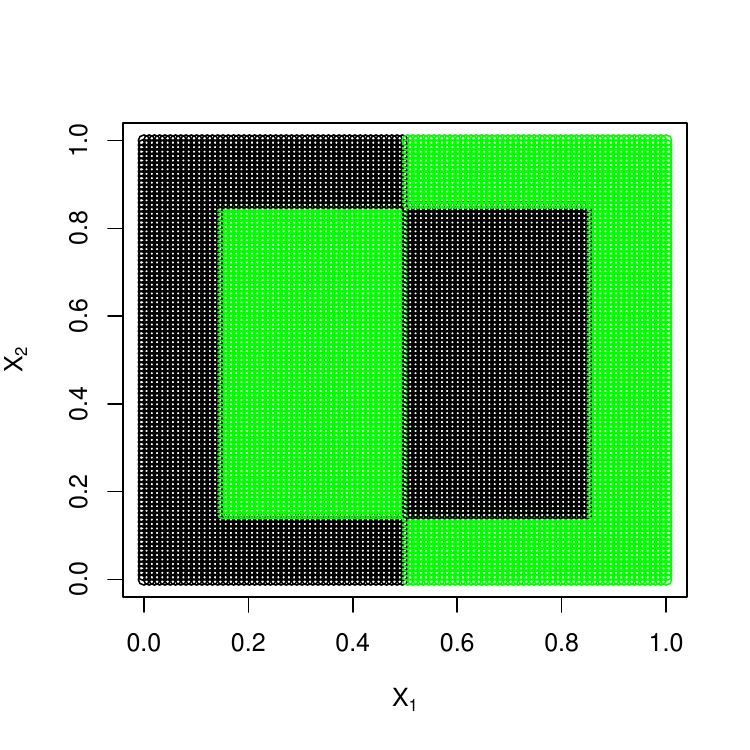}
\includegraphics[scale = 0.50,trim=5 10 20 25]{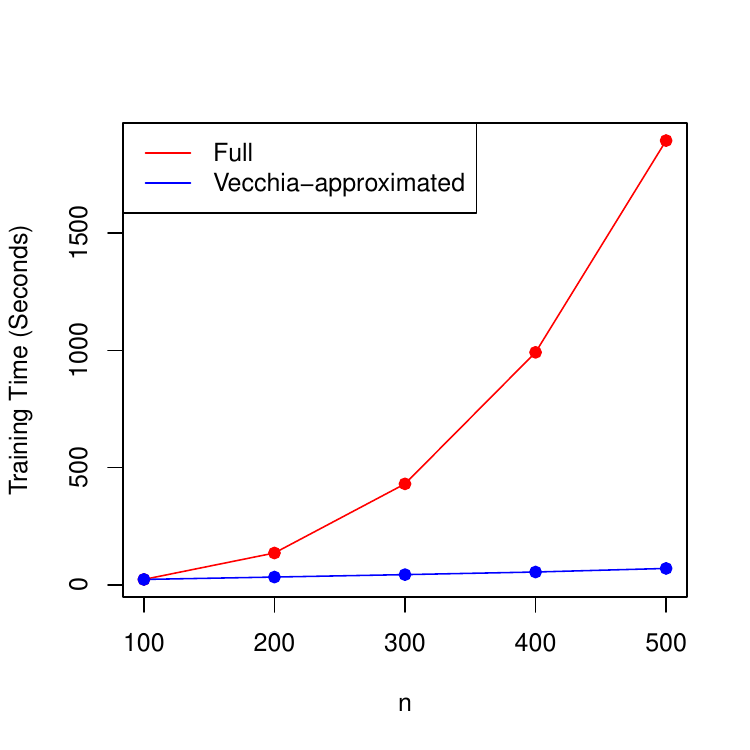}
\includegraphics[scale = 0.50,trim=5 10 30 25]{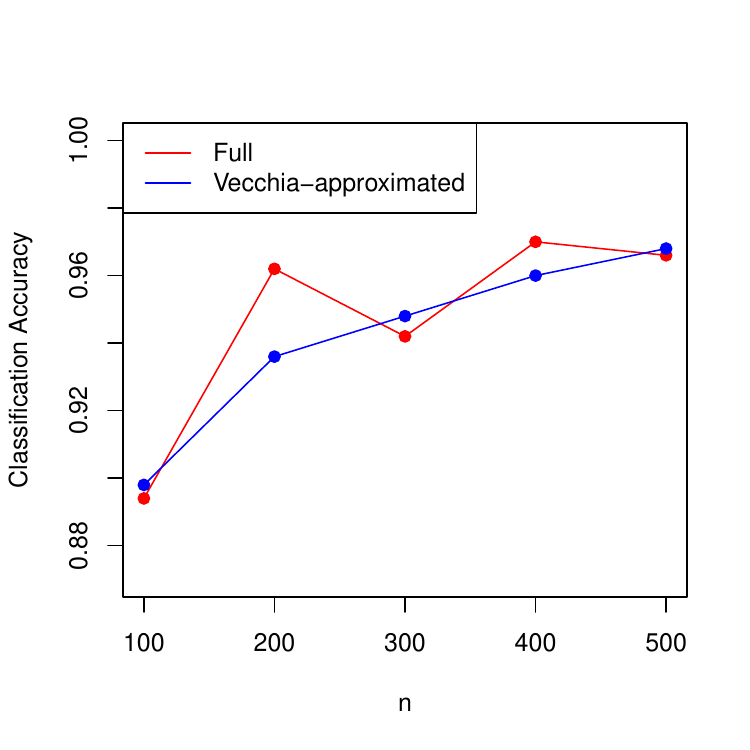}
\caption{{\em Left:} 2d ``box'' example. {\em Middle:}
run-times for ``full'' and Vecchia-approximated ($m=25$) GPC training. 
{\em Right:} classification accuracy from various GPCs.}
\label{fig:box}
\end{figure}

To showcase the speed-up Vecchia provides to GPC training, consider the
simulated 2d ``box'' example \citep{broderick2011classification} shown in the
left panel of Figure \ref{fig:box}.  
The middle panel compares the run-time for training a GPC to data from this
surface using $T=5{,}000$ MCMC iterations, with and without Vecchia.  For
both, we fix $\theta = 1$ (later estimated in Section \ref{sec:latent}) and
$\tau^2 = 25$ (discussed in more detail in Section \ref{sec:imp_details}) to
focus our comparison on the time required for latent $Z$ sampling.  Observe
that as the training size $n$ increases, the ``full'' GPC takes drastically
longer, while the Vecchia version only sees a slight bump in run-time. The
right panel compares predictive accuracy from 10 different LHS design
constructions on a separate LHS testing set of size $n' = 500$. Details are in
Section \ref{sec:latent}; the main takeaway here is that the
Vecchia-approximated GPC performs similarly to the full one.

\subsection{Full posterior sampling and prediction}
\label{sec:latent}

Here we complete the description of our Bayesian inferential framework for all
unknown quantities, connecting Sections \ref{sec:ess}--\ref{sec:vecchia} with
the ultimate goal of furnishing predictions (\ref{eq:gpcpost}). Algorithm
\ref{alg:gibbs_binary} outlines the ESS/Metropolis-within-Gibbs procedure we
use for MC integration over latent $Z$ and lengthscale $\theta$. 
Recall, $\theta$ is tacitly involved in $\Sigma(X)$, and therefore 
also in $U$ via Eq.~(\ref{eq:u_mat}).

\medskip
\begin{algorithm}[ht!]      
\DontPrintSemicolon
Initialize $\theta^{(1)}, Z^{(1)}$\;
    \For{$t = 2, \dots, T$}{
    $\theta^{(t)} \sim \pi(\theta\mid X, Z^{(t-1)})$ 
    \tcp*{MH via $\mathcal{L}(Z \mid X, \theta)$}
    $Z^{(t)} \sim \pi(Z \mid X,Y,\theta^{(t)})$
    \tcp*{ESS via $\mathcal{L}(Z \mid X, \theta)$, $\mathcal{L}(Y \mid Z)$}}
\caption{Gibbs sampling procedure for GPC estimation.}
\label{alg:gibbs_binary}
\end{algorithm}

Posterior sampling of $Z$ involves Vecchia-approximated MVN proposals within
ESS (Alg.~\ref{alg:ess}), where acceptance is based on the Bernoulli
likelihood (\ref{eq:class_like}).  Posterior sampling of $\theta$ employs
Metropolis Hastings, where acceptance probabilities utilize the MVN likelihood
of Eq.~(\ref{eq:like}), again with $Z$ in place of $Y$.  This calculation
requires an $n\times n$ inverse and determinant. Usually these are furnished
by a single Cholesky decomposition, which means the Vecchia approximation is
again handy here. In fact, speedy MVN likelihood calculations are the most
common application of Vecchia in the literature
\citep[e.g.,][]{katzfuss2020,datta2016hierarchical}. 
The details are provided here for completeness:
\begin{align*}
\mathcal{L}(Z \mid X) &\propto |\Sigma(X)|^{-\frac{1}{2}}\exp\left\{-\frac{1}{2}Z^\top\Sigma(X)^{-1}Z\right\} \\
    &\approx \sum_{i=1}^n U^{ii} \cdot \exp\left\{-\frac{1}{2}Z^\top UU^\top Z\right\}.
\end{align*}
Given $X$ and $\theta^{(t)}$, $U$ might be more aptly notated as $U^{(t)}(X)$,
but we opt to keep it simple. Since $U$ uses $\theta^{(t)}$, it must be
reconstructed if a new $\theta$-value is accepted.  As with MVN proposals
in Section \ref{sec:vecchia}, the computational order for likelihood
evaluation under Vecchia is $\mathcal{O}(n m^3)$. Additional details on our
prior and proposal mechanism for $\theta$, along with choices of initial
values $(\theta^{(1)}, Z^{(1)})$, are reserved for Section
\ref{sec:imp_details}.

We pivot now to prediction.  Suppose we have a collection of testing inputs
$\mathcal{X}\in\mathbb{R}^{n' \times d}$ where we wish to assess
$\mathcal{Y}(\mathcal{X}) \mid X, Y$. In the context of a single predictive
location $x\in \mathcal{X}$, Eq.~(\ref{eq:gpcpost}) explains that after we
integrate over $Z$ [Section \ref{sec:vecchia} and Alg.~\ref{alg:gibbs_binary}], 
all that remains is to generate $z(x) \mid Z, X$ and
apply the sigmoid $\sigma(z(x))$.  Here we flesh out the details of prediction
vectorized over all $\mathcal{Z} \equiv \mathcal{Z}(\mathcal{X})$.  Given a
sample of $Z^{(t)}$-values, we may draw $\mathcal{Z}^{(t)}$-values via
Eq.~(\ref{eq:pred}), again with $Z^{(t)}$ and $\mathcal{Z}^{(t)}$ rather than
$Y$ and $\mathcal{Y}$, which may be computationally fraught without
approximation.  As a workaround, we appropriate \citet{Sauer2022}'s simplified
Vecchia scheme, which may be described as follows.

First form $X^{\text{stack}} = [X;\mathcal{X}]$ by stacking one after the other, 
row-wise, letting $i=1,\dots,n,n+1,\dots,n+n'$ index the rows.  Note, testing
locations must be indexed {\em after} training ones.
Next extend conditioning sets $c(i)$ for $i=n+1,\dots,n+n'$ to include any 
indices $j < i$, meaning testing locations might condition on other testing
locations.  Then, apply Eq.~(\ref{eq:u_mat}) with the combined $X^\text{stack}$ 
in place of $X$ to obtain what we refer to as $U^\text{stack}$.  This
matrix is still sparse and upper triangular, with at most $m$-many non-zero 
off-diagonal elements in each column.  We may partition $U^\text{stack}$ 
as follows,
\[
U^\text{stack} = 
\begin{bmatrix} U_X & U_{X,\mathcal{X}} \\ 0 & U_\mathcal{X}\end{bmatrix},
\]
where $U_X$ is the original upper triangular matrix formed from Eq.~(\ref{eq:u_mat})
with $X$ alone.  Then, one may derive the following predictive equations:
\begin{align}
\mathcal{Z}^{(t)} \mid X,Y,Z^{(t)},\theta^{(t)} \sim 
\mathcal{N}_{n'}(\mu^*,\Sigma^*) && & \mbox{where} & 
\mu^* &= -(U^\top_\mathcal{X})^{-1}U^\top_{X,\mathcal{X}}Z^{(t)} \quad \mbox{and} \quad 
\Sigma^* = \left(U_\mathcal{X}U_\mathcal{X}^\top\right)^{-1}. 
\end{align} 
Again, note both $\mu^*$ and $\Sigma^*$ tacitly depend on $\theta^{(t)}$ via
$U_\mathcal{X}$ and $U_{X,\mathcal{X}}$. Sampling from this MVN can be
computationally intensive for $d>1$, even with Vecchia applied. One way to
further speed up prediction is to draw predictive samples at individual
entries in $\mathcal{X}$ via their univariate normals. While pointwise
prediction neglects covariances across out-of-sample observations, we have
found predictions are nonetheless very similar. We therefore utilize this
shortcut in all of our experiments moving forward.

Samples $\mathcal{Z}^{(t)}$ collected over $t=1,\dots,T$ may 
be converted into assessments of $p_y(\mathcal{X})$, completing our 
vectorized approximation to Eq.~(\ref{eq:gpcpost}), through the
laws of total expectation and variance:
\begin{equation}
\begin{split}
\hat{\mu}_y(\mathcal{X}) &= \dfrac{1}{T}\sum_{t=1}^T\sigma\left(\mathcal{Z}^{(t)}\right)\\
\hat{\sigma}^2_y(\mathcal{X}) &= \dfrac{1}{T-1}\sum_{t=1}^T\left(\sigma
\left(\mathcal{Z}^{(t)}\right) - \hat{\mu}_y(\mathcal{X})\right)^\top
\left(\sigma\left(\mathcal{Z}^{(t)}\right) - \hat{\mu}_y(\mathcal{X})\right)
 + \dfrac{1}{T}\sum_{t=1}^T
\sigma\left(\mathcal{Z}^{(t)}\right)\left(1 - \sigma\left(\mathcal{Z}^{(t)}\right)\right).
\end{split}
\label{eq:post_pred}
\end{equation}
The first term in Eq.~(\ref{eq:post_pred}) for $\hat{\sigma}^2_y(\mathcal{X})$ represents 
uncertainty in the model's estimated probability of success, while the second stems 
from inherent variability contained in the random Bernoulli draw which produced the binary 
output. The latter component increases as estimated probabilities approach $0.5$, where drawing 
$y(x)$ becomes essentially a coin flip. A classification rule can be formed from any threshold on 
$\hat{\mu}_y$; however, in our own work (such as the right panel of Figure \ref{fig:box}), 
we stick with the 50:50 rule: $\hat{y}(x_i) = \mathbbm{1}\left(\hat{\mu}_y(x_i) \geq 0.5\right)$.

\begin{figure}[ht!]
\centering
\includegraphics[scale=0.6,trim=0 10 0 40,clip=TRUE]{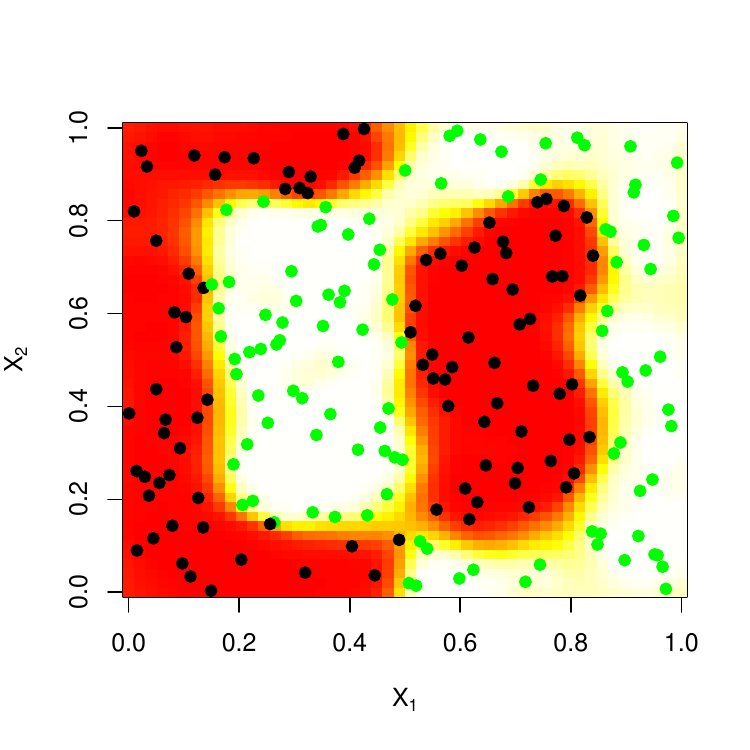}
\includegraphics[scale=0.6,trim=50 10 0 40,clip=TRUE]{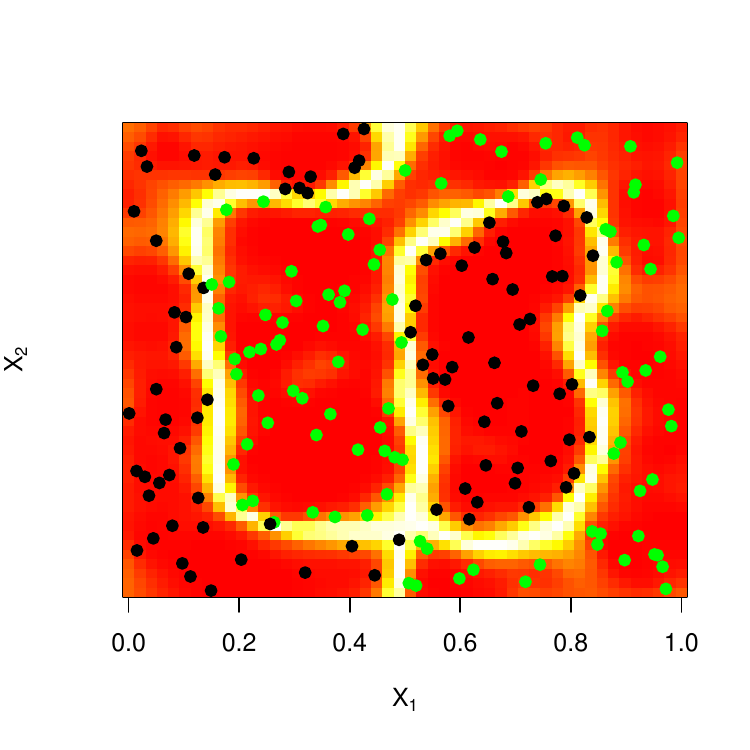}
\caption{Posterior mean (\textit{left}) and variance (\textit{right}) of $p_y(\mathcal{X})$
for the 2d ``box'' example [Figure \ref{fig:box}] produced by a Vecchia-approximated 
GPC. Brighter colors indicate larger values.}
\label{fig:box_fit}
\end{figure}

To continue the illustration, consider Figure \ref{fig:box_fit}, which
provides views of $\hat{\mu}_y(\mathcal{X})$ and
$\hat{\sigma}^2_y(\mathcal{X})$ from Eq.~(\ref{eq:post_pred}) in its left and
right panels, respectively.  As in Figure \ref{fig:box} we use $m=25$, but for
this illustration we fix $n=200$ via LHS and use a regular predictive grid for
$\mathcal{X}$ with a total of $n'=10{,}000$ rows. Notice the mean predictions
in the left panel capture the general form of the box shape in the response
surface. The uncertainty shown in the right panel highlights boundary regions
dividing class labels. This makes intuitive sense; uncertainty is highest in
locations where the response transitions from success to failure.

\section{Implementation and empirical results}
\label{sec:implement}

Here we provide implementation details for our proposed GPC model, and assess
its performance in both simulated examples and the motivating binary black
hole example introduced in Section \ref{sec:1}.

\subsection{Implementation details}
\label{sec:imp_details}

Our implementation allows for either the squared-exponential (\ref{eq:sigma})
or Matérn $\nu = 5/2$ kernels for $\Sigma(X)$,
with the latter performing slightly better in our comparisons. Both require
specification of lengthscale $\theta$ and scale $\tau^2$. We fix $\tau^2$
following a simple rule (discussed momentarily) and sample $\theta$ via MCMC
alternating with $Z$ in a Gibbs-like fashion [Alg.~\ref{alg:gibbs_binary}].
For all empirical work we take $\theta
\sim \mathrm{Gamma}(1.5, 2.6)$ providing a prior spread appropriate for inputs
coded, marginally, to the unit interval. Metropolis proposals follow
$\theta^{(t)}\sim
\text{Unif}\left[u\theta^{(t-1)},\dfrac{1}{u}\theta^{(t-1)}\right]$.
Throughout we use $u = 2/3$, i.e., proposing $\approx 33\%$ above or below the
previous value, since that consistently yields a reasonable acceptance rate.
Chains are initialized with $\theta^{(1)} = 0.1$ and $z_i^{(1)} = 2\tau$ or
$-2\tau$ depending on if $y_i = 1$ or $0$, respectively. All chains are run
for $10{,}000$ iterations, discarding the first $1{,}000$ as burn-in and
thinning by a factor of $10$ for a total of $T=900$.

Although this choice of $Z^{(1)}$ gives decent initial probabilities
$\sigma(Z)$ associated with $Y$, it is not a smooth (or stationary) function
of inputs $X$, and therefore presents a poor initialization of a latent GP.
When $n$ is small, burn-in to a smooth GP realization is quick, but for larger
$n$ we find it is helpful to introduce a nugget \citep{gramacy2012cases}
hyperparameter into $\Sigma$ which captures deviations from a
smooth/stationary process as noise \citep{binois2018practical}.  We only do
this for the burn-in portion of our sampling, taking a zero nugget (i.e., no
nugget and no noise) afterwards for posterior sampling.  During burn-in, we
additionally find it helpful to utilize a prior on the nugget that
concentrates on zero as sampling iteration $t$ progresses, i.e.,
$\mathrm{Gamma}(1, 10t)$.  The effect of a (non-zero) nugget parameter has
long remained one of speculation in the GPC literature, going back to
\citet{bernardo1998regression}.  Our consideration here is somewhat different,
focusing on the nugget's effect on what we would otherwise prefer to be smooth
latent quantities, considering our deterministic surrogate modeling context.
\citeauthor{binois2018practical} remark that in their setting, which is
similar to but not identical to ours, the maximum {\em a posterior} nugget is
zero, so there is little harm in concentrating it out artificially, as we do
during burn-in via the prior.

Our Vecchia implementation leverages a parallelized construction of $U_X$,
$U_\mathcal{X}$ and $U_{X,\mathcal{X}}$, independently distributing rows over
multiple cores via {\tt OpenMP} and {\tt RcppArmadillo}
\citep{eddelbuettel2014rcpparmadillo} and as described by \cite{Sauer2022}.
Sparse matrix storage and manipulation is furnished via the {\tt Matrix}
library \citep{bates2010matrix}. We use random indexing and nearest
neighbors for $c(i)$ as is common in the literature
\citep[e.g.,][]{stein2004, datta2016hierarchical, katzfuss2020}.

We compare the predictive performance of our full (where possible) and
Vecchia-approximated GPC predictions to the following benchmark comparators:
\begin{itemize}
\item Doubly-Stochastic Variational Inference \citep[DSVI,][]{salimbeni2017},
implemented in the {\sf Python} library {\tt GPFlux}
\citep{dutordoir2021gpflux}
\item Scalable Variational Gaussian Process
\citep[SVGP,][]{hensman2015scalable}, implemented in the {\sf Python} library
{\tt GPflow} \citep{GPflow2017}.
\end{itemize}
In each of ten repeated MC instances, we randomly generate a training and
testing partition of sizes $n$ and $n'$, respectively, fit the models on the
training data, and then use them to predict out-of-sample on the testing set.
Throughout, we fix $n' = 1{,}000$ while varying $n$ over a range that is
test-problem dependent.  The comparisons coming shortly additionally show a
deep (``DGPC'') competitor which can be ignored for now and will be discussed,
along with additional analysis and commentary, in Section \ref{sec:deepgp}.

Performance metrics include correct classification rate (CR, higher is better) and 
logarathmic score \citep[LS, higher is better;][Table 1]{gneiting2007strictly}. 
Let $y_i$ denote the true
class label associated with testing input $x_i \in
\mathcal{X}$, $i=1,\dots, n'$, collected as $\mathcal{Y}$. Then, let
$\hat{y}_i \equiv
\hat{y}(x_i) \mid X, Y$ denote a predicted label given $n$ training examples
$(X,Y)$, collected as $\hat{\mathcal{Y}}$.  Similarly let $\hat{p}_i \equiv
p_y(x_i) \equiv \mathbb{P}(Y(x_i) = 1 \mid X,Y)$ denote the predicted probability $y_i \in
\mathcal{Y}$ is 1, and collect these as $\hat{\mathcal{P}}$. Then CR and
LS are defined as follows:
\begin{align}
\mathrm{CR}(\mathcal{Y}, \hat{\mathcal{Y}}) &=  \label{eq:mets}
\frac{1}{n'}\sum_{i=1}^{n'} \mathbbm{1}\left(y_i=\hat{y_i}\right) &
\mathrm{LS}(\mathcal{Y}, \hat{\mathcal{P}}) &= 
-\frac{1}{n'}\sum_{i=1}^{n'}\left[y_i\log(\hat{p}_i)+(1-y_i)\log\left(1-\hat{p}_i\right)\right].
\end{align}
When evaluating our {\sf Python} comparators on these metrics, we
simply take $\hat{\mathcal{P}}$ from the software output directly. For our own
methods, we use posterior sampling as described earlier and take
$\hat{\mathcal{P}} =
\hat{\mu}_y(\mathcal{X})$ following Eq.~(\ref{eq:post_pred}).  For all
methods we convert $\hat{\mathcal{P}}$ to $\hat{\mathcal{Y}}$ using a
threshold of $\hat{\mu}_y(\mathcal{X}) \geq 0.5$.

\subsubsection*{Choosing an appropriate scale for the reference process}

An inverse-gamma (IG) prior on $\tau^2$ is semi-conjugate, meaning conditional 
posterior inference may be facilitated by analytic integration without
MCMC sampling. An (improper) ``reference'' prior $p(\tau^2)
\propto \dfrac{1}{\tau^2}$, which is mathematically IG$(0,0)$, is similarly
analytic and thus a common choice in regression GP settings \citep[Chapter
5]{gramacy2020surrogates}. However, we have found that, in our classification
setting, posterior integration over $\tau^2$ does not work well because the
responses (which are just $0$s and $1$s) do not provide a strong sense of
``scale'' for the latent $Z$ process via the Bernoulli likelihood.  A
practical consequence of this is that posterior draws for $\tau^2$ can be
wildly large under a reference prior, translating to very large/small
probabilities $\sigma(Z)$ under the logit transform, and likewise ones that
are too close to $\sigma(Z) = 0.5$ under stronger priors even when
prediction ought to be more confident.

\begin{table}[ht!]
\centering
\begin{tabular}{r|rrrrr}
$\tau^2$ & $2^0$ & $2^1$ & $2^2$ & $2^3$ & $2^4$ \\
\hline
$\sigma(2^\tau)$ & 0.881 & 0.944 & 0.982 & 0.997 & 1.000 \\
$\sigma(-2^\tau)$ & 0.119 & 0.056 & 0.018 & 0.003 & 0.000 \\
\end{tabular}
\caption{\label{t:tau2} Impact of the choice of $\tau^2$ on tail
classification probabilities.}
\end{table}
 
To provide more detail, and ultimately aid in prescribing an appropriate,
data-dependent setting for the latent scale, consider the values provided in
Table \ref{t:tau2}.  As $\tau^2$ grows from $\tau^2 = 1$ to $16$ the
table shows the range of low and high probabilities that could, {\em a
priori}, be realized from $Z$-values out at 95\% interval boundaries. For
example, when $\tau^2 = 2^0 = 1$ most $Z$s are ({\em a priori}) within
$[-2,2]$ which means $p\in [0.119,0.881]$ with probability 0.95.  Such a range
may be appropriate with a small training data set, like $n \leq 10$. Suppose
you believed {\em a priori} that you {\em could} observe at least one
observation in each class, anywhere in the input space.  When $n = 10$ that
means $p(x) \in [0.1, 0.9]$, i.e., a similar range as in the table for $\tau^2
= 1$.  Such a prior could be said to follow a unit-information principle
\citep{jin2021unit}.  Now imagine a balanced training data set of $n=2{,}000$
points (1000 in each class) were labels are geographically separated from one
another.  In that case $p(x) = 999/1{,}000$ or $1/1{,}000$ is reasonable for any
particular $x$ regardless of its location in the input space.  A prior for
$Z$ via $\tau^2=1$ would be a poor match, shrinking $p(x)$ towards
$[0.119,0.881]$ leading to under-confidence and lower scores (\ref{eq:mets})
out-of-sample. Looking at Table \ref{t:tau2}, a better match to that scenario
would be $\tau^2 = 2^3$ so that $p\in [0.003,0.997]$.

We wish to operationalize this idea into a scheme that provides an automatic
determination for $\tau^2$ in a data-dependent way, but crucially avoids
``double-dipping'' from a Bayesian statistical perspective.
To this end, we introduce a concept of a training data point's degree of
{\em insulation}: $\omega_i \equiv \omega(x_i, y_i)$, which is the size of the
(nearest) neighborhood of $x_i$ in the training data set $X$ such that all of
$x_i$'s neighbors in $X$ agree in label with $y_i$.  Insulation is thereby
measuring a degree of local homogeneity.  Specifically, let $D =
\{d_{ij}\}_{i,j=1}^{n}$ denote the $n\times n$ pairwise Euclidean distance
matrix of $X$, and define
\begin{align}
\omega_i &= \left|\left\{x_j \in X^{(-i)} : d_{ij} < \min_{k\neq
i}\left\{d_{ik} : y_k\neq y_i\right\}\right\}\right|. \label{eq:insul}
\end{align}

Now let $\omega^{\max} = \max_{i = 1,\dots,n} \omega_i$ denote the degree of
the most insulated training data point.  The unit-information-prior idea above
suggests that a sensible setting for $\tau^2$ could be derived by anchoring to
$\omega^{\max} / (\omega^{\max} + 1)$ and its complement $1 /
(\omega^{\max} + 1)$.  Here we introduce a tuning parameter $\epsilon$ to
entertain an $\epsilon$-information prior via $p^{\max}_\epsilon =
\omega^{\max} / (\omega^{\max} + \epsilon)$.  Then, invert $p^{\max}_\epsilon$
with $\sigma(\cdot)$ to determine the latent $z$-value; under a logit
transform, that is $z^{\max}_\epsilon =
\log\left(p^{\max}_\epsilon/(1 - p^{\max}_\epsilon)\right)$. Finally,
determine $\tau^2$ so that $z^{\max}_\epsilon$ is two standard deviations from
the mean ($z^{\max}_\epsilon = 2\tau$) and solve: $\tau^2_\epsilon =
\left(z^{\max}_\epsilon/2\right)^2$.

\begin{figure}[ht!]
\centering
\includegraphics[scale=0.65, trim=0 10 0 40,clip=TRUE]{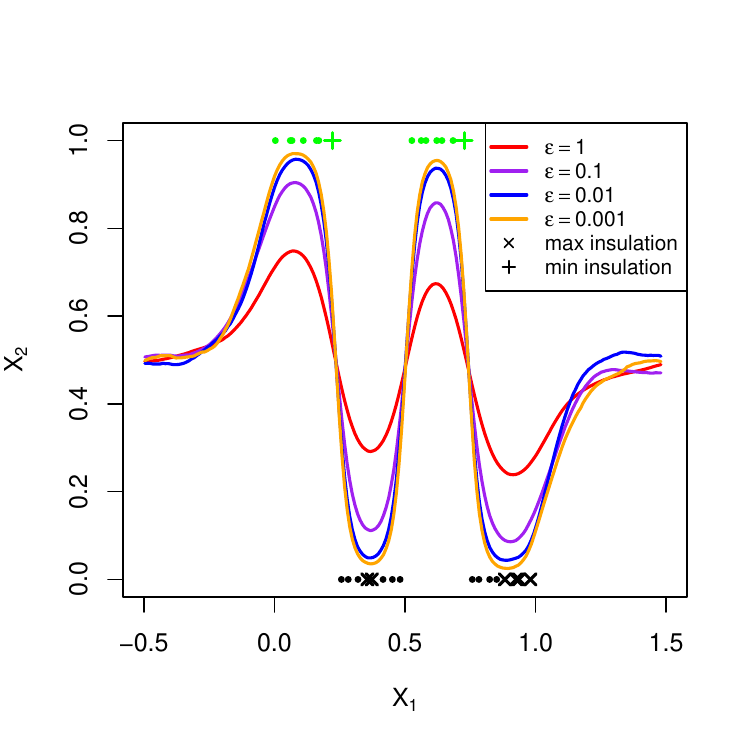}
\includegraphics[scale=0.65, trim=0 10 0 40,clip=TRUE]{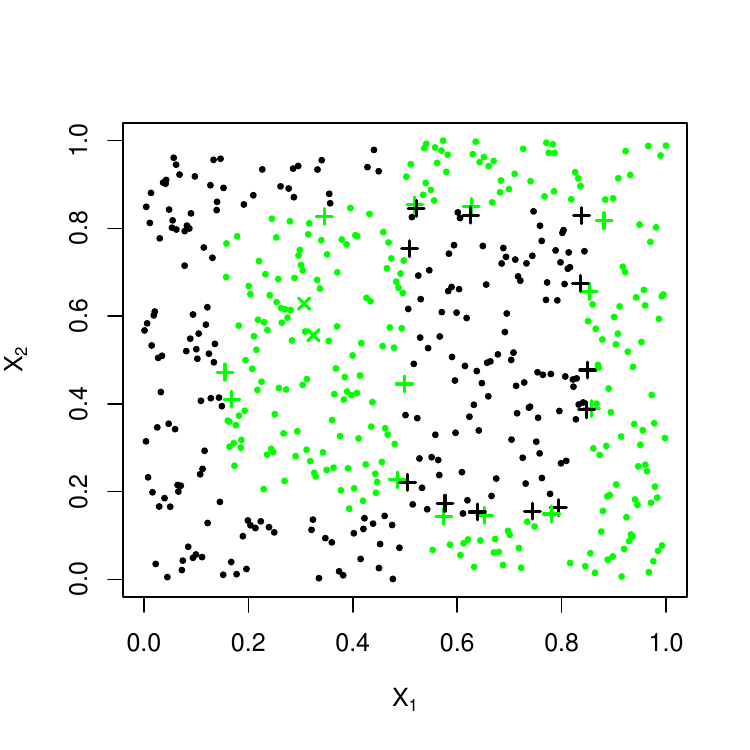}
\caption{{\em Left:} Posterior mean $\mu(x)$ for various $\epsilon$ values 
specifying $\tau^2$  on the 1d sin example [Figure \ref{fig:1d_ex}]. $\times$'s
denote observations with maximum insulation ($\omega^{\max}=7$). {\em Right:}
$500$ randomly sampled points from the 2d box example [Figure \ref{fig:box}],
with $\omega^{\max} = 55$. For reference, $+$'s indicate minimum insulation.}
\label{fig:scale_ex}
\end{figure}

Two examples are provided in Figure \ref{fig:scale_ex}, via our earlier 1d
[Section \ref{sec:class_ex}/Figure \ref{fig:1d_ex}] and 2d [Section
\ref{sec:latent}/Figure \ref{fig:box_fit}] illustrations. The most-insulated 
points are indicated with $\times$'s, where $\omega^{\max}$ is $7$
for the first example and $55$ for the second.  A choice of $\epsilon$
determines $p^{\max}_\epsilon$ and that, in turn, affects the posterior mean
prediction from our GPC model, as indicated in the left panel for the 1d
example. A unit-information prior ($\epsilon = 1$) would be sensible in many
modeling contexts; however, that is a poor choice for this example, and would
lead to a washed-out predictive surface (in red).  One thing that is special
in our context of surrogate modeling for computer simulation experiments is
that the data-generating mechanism is deterministic.  Thus,
for well-insulated data points, the probabilities won't change quickly
as we move away from such locations.  In other words, we can get away with a
much smaller $\epsilon$.

One option is to dial in $\hat{\epsilon}$ via cross-validation (CV).  Although
this could require substantial work, recognizing additional
structure offers some simplification. Observe there is a monotonic
relationship between $\tau_\epsilon^2$ and $p^{\max}_\epsilon$.  As $\epsilon
\rightarrow 0$, say beginning from the unit information prior $\epsilon = 1$,
we get larger $\tau_\epsilon^2$ yielding larger $p^{\max}_\epsilon$.  This is
shown in the left panel of Figure \ref{fig:scale_ex} via
predictive surfaces.  Also note that, for fixed $\epsilon$, we would generally
have larger $\tau^2_\epsilon$ and $p^{\max}_\epsilon$ as $n$ increases,
since that would generally result in a greater degree of insulation
(\ref{eq:insul}) and thus larger $\omega^{\max}$.  But details would still depend
on the stochasticity of the data-generating mechanism.  For deterministic
responses this relationship is monotonic, so a deterministic CV (like
leave-one-out) could proceed via bisection search, which would dramatically reduce
the computational effort.  However, we have found even that to be overkill. In
all of our examples going forward we fix $\varepsilon = 0.001$.  For Figure
\ref{fig:scale_ex} this results in $\tau^2_\epsilon$ values of about $20$ and
$30$, respectively.  Accordingly, our prior allows our most insulated point to 
``reach'' a small/large probability at least as extreme as $1/1{,}000$ and 
$999/1{,}000$, possibly even more extreme when $n$ and $\omega^{\max}$ are 
large.  Note, this does not preclude learning something smaller via 
the posterior.

\subsection{Simulated examples}
\label{sec:sim_ex}

\paragraph{2d Box.} We first return to the two-dimensional ``box'' example shown in Figure \ref{fig:box}. Due to the simplicity of this example, the results from our Monte Carlo experiment are not particularly noteworthy. We found all methods performed well for all training sizes, apart from DSVI, which achieved much lower log scores than the rest. Results from our experiment can be seen in Appendix \ref{appendix:box} for those who are curious.

\begin{figure}[ht!]
\centering
\begin{tabular}{cc}
\begin{minipage}{6cm}
\includegraphics[scale=0.33,trim=0 0 0 0]{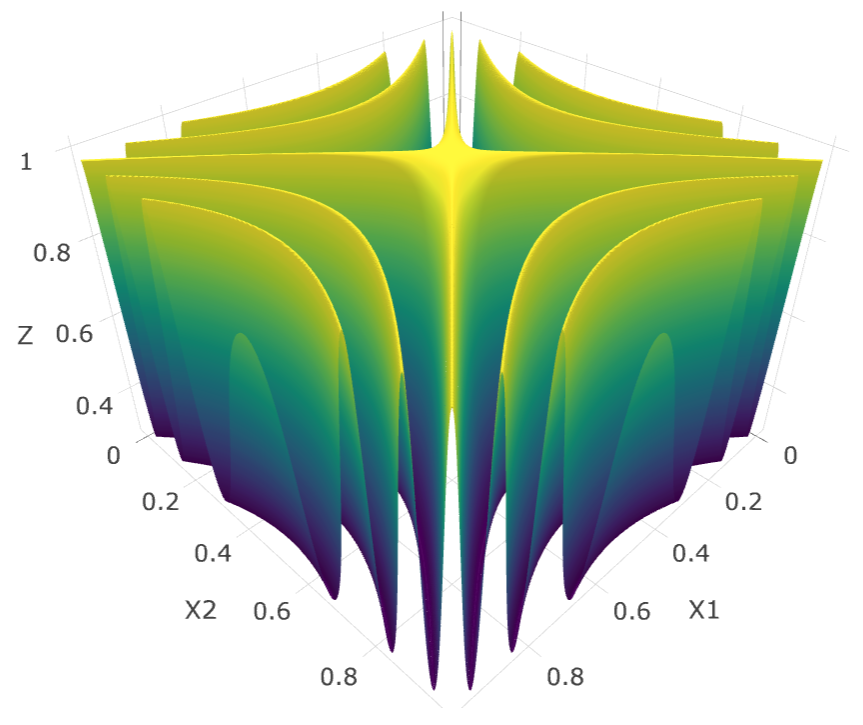}
\end{minipage}
\begin{minipage}{5.4cm}
\includegraphics[scale=0.5,trim=0 10 0 40]{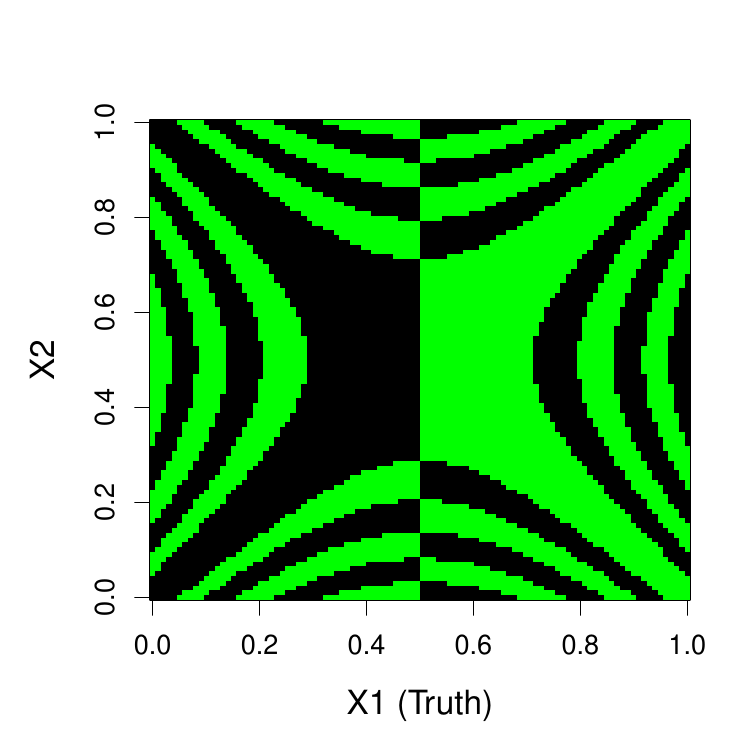}
\end{minipage}\\
\begin{minipage}{6.6cm}
\includegraphics[scale=0.5,trim=-20 10 0 40]{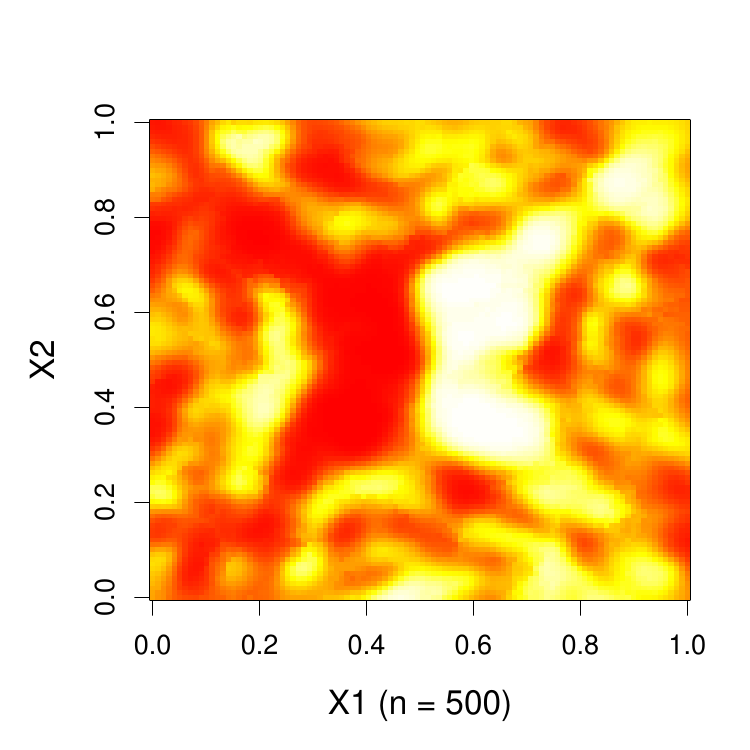} 
\end{minipage}
\begin{minipage}{6cm}
\includegraphics[scale=0.5,trim=-10 10 0 40]{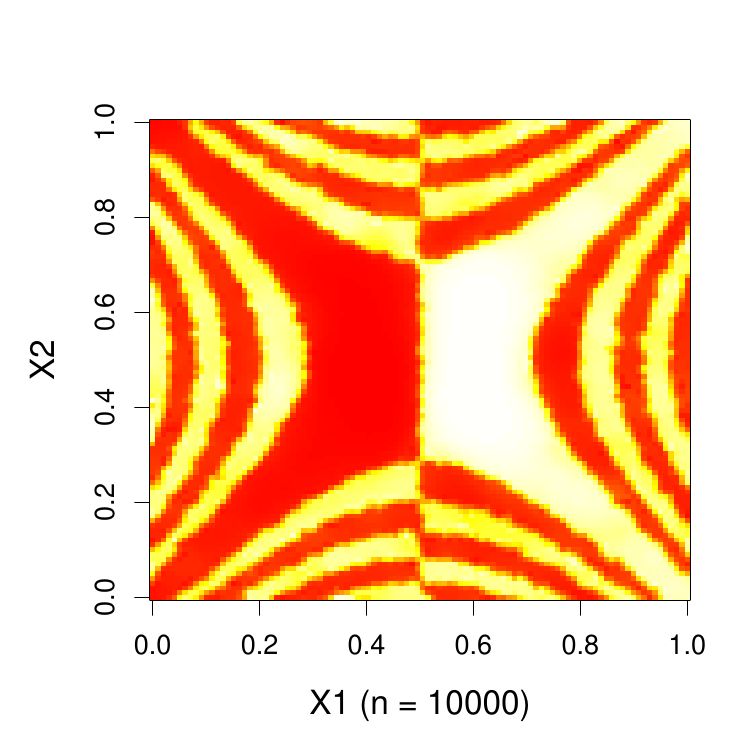}
\end{minipage}
\end{tabular}
\caption{{\em Top-left}: Schaffer no.~4 function on
$X\in[-2,2]^2$. {\em Top-right}: Binarized version colored by class membership
($\text{black} = 0$, $\text{green} = 1$), and scaled to input space
$X\in[0,1]^2$. {\em Bottom}: Vecchia-GPC posterior mean predictions with 
$n=500$ ({\em left}) and $n=10{,}000$ ({\em right}) LHS training points, where brighter colors indicate larger success probabilities.}
\label{fig:schaffer}
\end{figure}

\paragraph{2d Schaffer no.~4.} For a more complex 2d example, we follow
\cite{broderick2011classification} and construct a classification problem out
of a common regression example using components of partial derivatives. We
borrow the Schaffer no.~4 function from the Virtual Library of Simulation
Experiments \citep[VLSE;][]{simulationlib}, illustrated in the top-left panel
of Figure \ref{fig:schaffer}. The real-valued function is binarized using the
$x_1$-component of its gradient, where $y_i = 1$ if $\dfrac{\partial
z}{\partial x_{i1}} > 0$, and $y_i=0$ otherwise. The result is shown in the
top-right panel, with two colors indicating the two unique classes.
\begin{figure}[ht!]
\centering
\includegraphics[scale=0.5,trim=20 10 0 0]{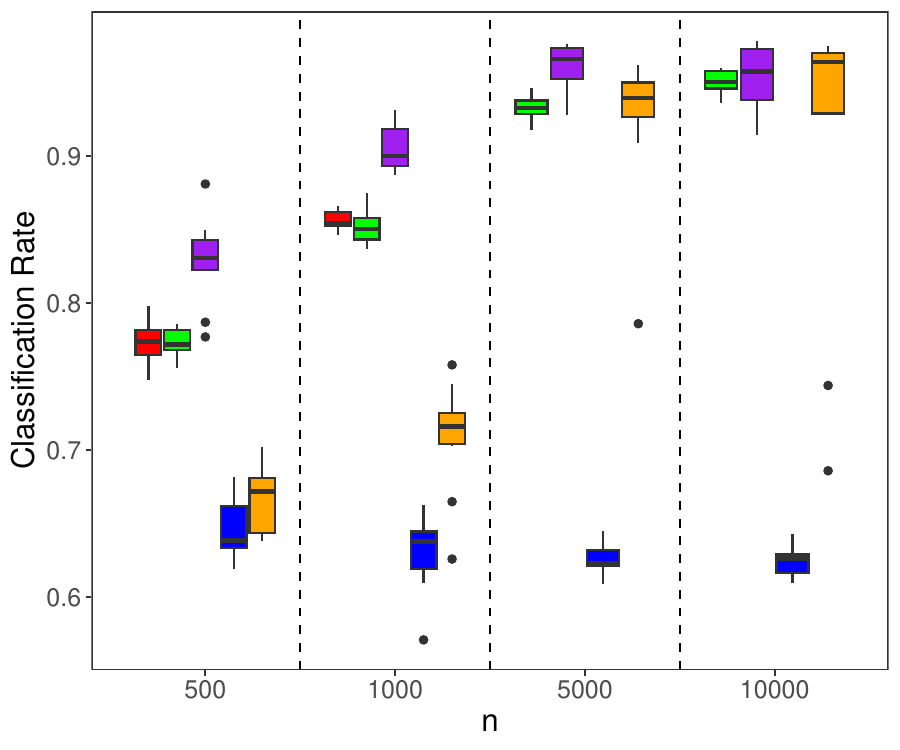}
\hspace{1cm}
\includegraphics[scale=0.5,trim=0 10 0 0]{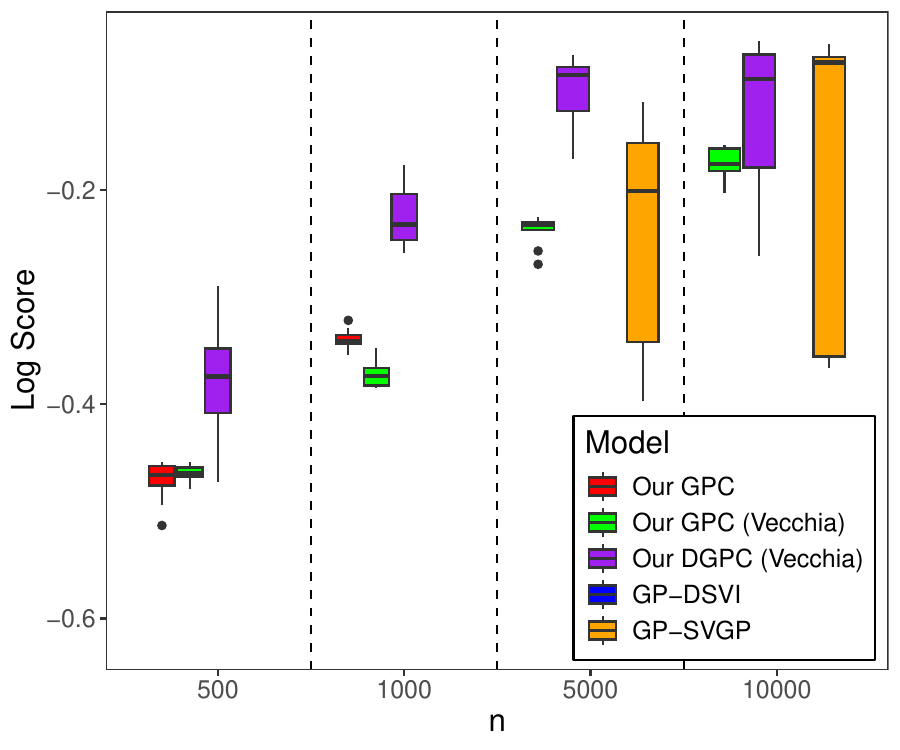}
\caption{Schaffer function no.~4 results.}
\label{fig:schaffer_results}
\end{figure}

Predictive performance is compared in Figure \ref{fig:schaffer_results}. For
all training sizes our Vecchia-GPC consistently outperforms alternatives
across both metrics. SVGP is only competitive for larger training sample
sizes. However, what is apparent is the large amount of variability in
performance exhibited by both SVGP and DSVI. Vecchia-GPC performance contains
considerably less variability, particularly in log score, which indicates more
effective UQ.  Obtaining good scores is of primary importance to us.   While
raw accuracy (CR) is intuitive, good UQ (via LS) is our main goal.

\paragraph{Six-dimensional ``G''.}  The ``G'' function
\citep{marrel2009calculations} has a jagged input-output dynamic that can be
evaluated in arbitrary dimension. We borrow the VLSE form with $X\in[0,1]^6$
and binarized real-valued output $Z$ as $y_i = 1$ if $z_i > 1$, and $y_i=0$
otherwise.
\begin{figure}[ht!]
\centering
\includegraphics[scale=0.5,trim=20 01 0 0]{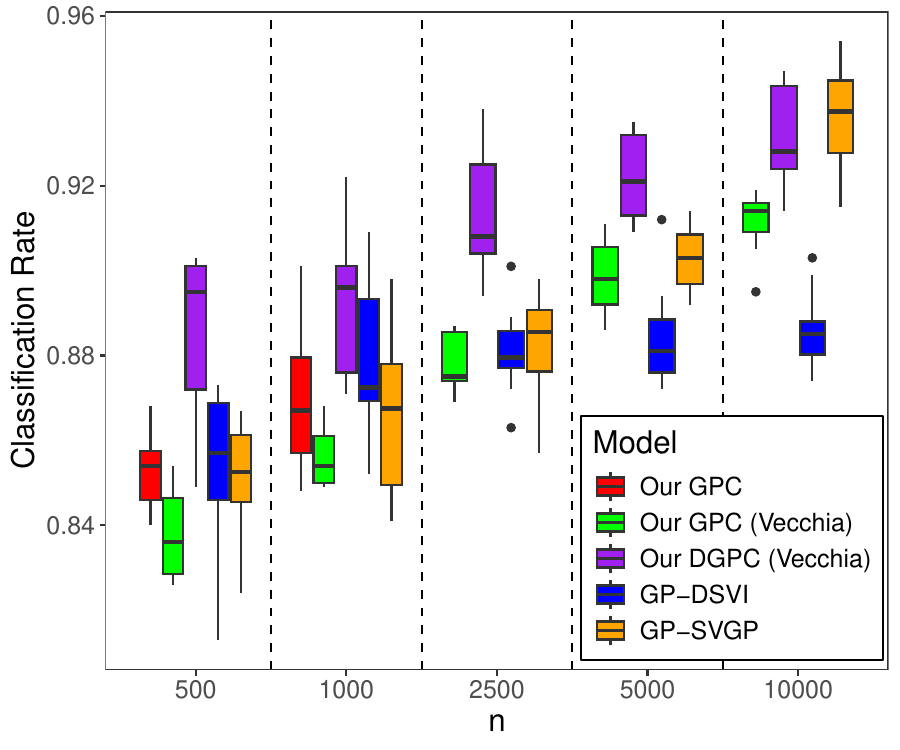}
\hspace{1cm}
\includegraphics[scale=0.5,trim=0 10 0 0]{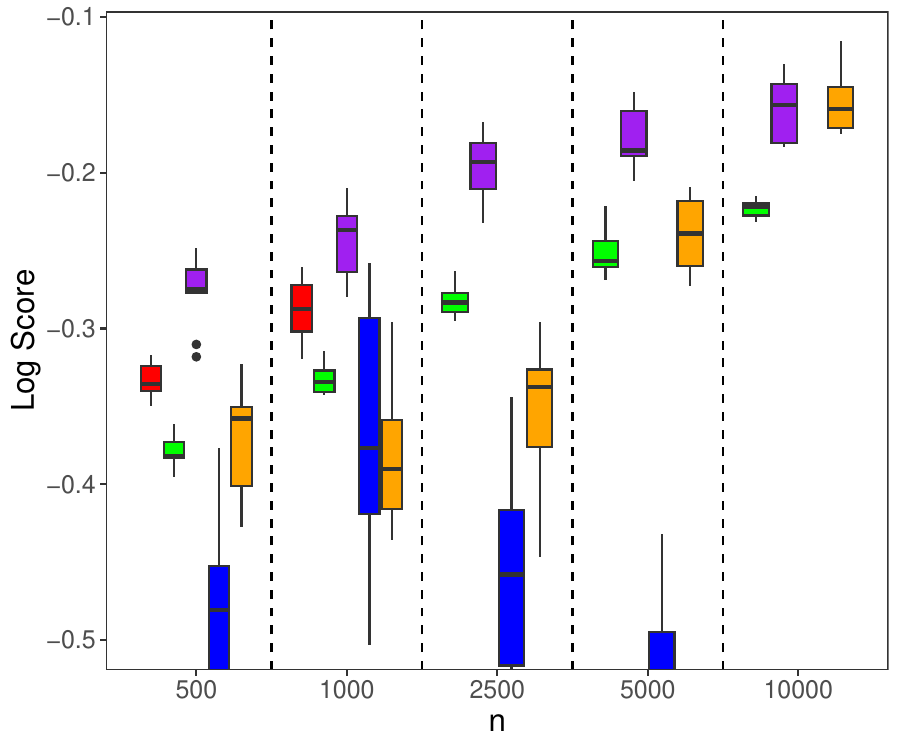}
\caption{``G'' function results.}
\label{fig:g6_results}
\end{figure}
Out-of-sample prediction results are summarized in Figure
\ref{fig:g6_results}. Although the CR of our Vecchia-GPC is slightly lower
than that of SVGP, it consistently has the highest LS across all MC iterations
for smaller training sizes and exhibits less variability. SVGP only achieves
higher LS when $n=10,000$, and seems to perform inconsistently with less data.
Our Vecchia-approximated approach doesn't perform appreciably different than
our full GPC for smaller $n$. 

\subsection{Binary black hole formation example}
\label{sec:bbh}

The COMPAS model \citep{compas2022} is used to study the formation of binary
black holes \citep[BBHs;][]{rauf2023exploring}. Eleven characteristics of two
celestial bodies are given as inputs, including their respective masses and
orbital separation, the magnitude and direction of the supernova natal kick
vector, and others. The mass of the BBH formed -- known as the ``chirp mass''
-- is returned as output. Since BBH formation is rare (around one in ten
   thousand), most input configurations result ``{\tt NA}'' instead of a valid
   chirp mass. So the response is of mixed type: binary (whether a BBH formed
   or not) and numeric (if one formed, what is its mass). More formally, if $x
   \in [0,1]^{11}$ represents coded inputs, and $m(x)$ is the output COMPAS,
   then $y(x) =
\mathbb{I}(m(x) \in \mathbb{R}^+)$ indicates BBH formation.
\begin{figure}[ht!]
\centering
\includegraphics[scale=0.5,trim=-10 10 0 40]{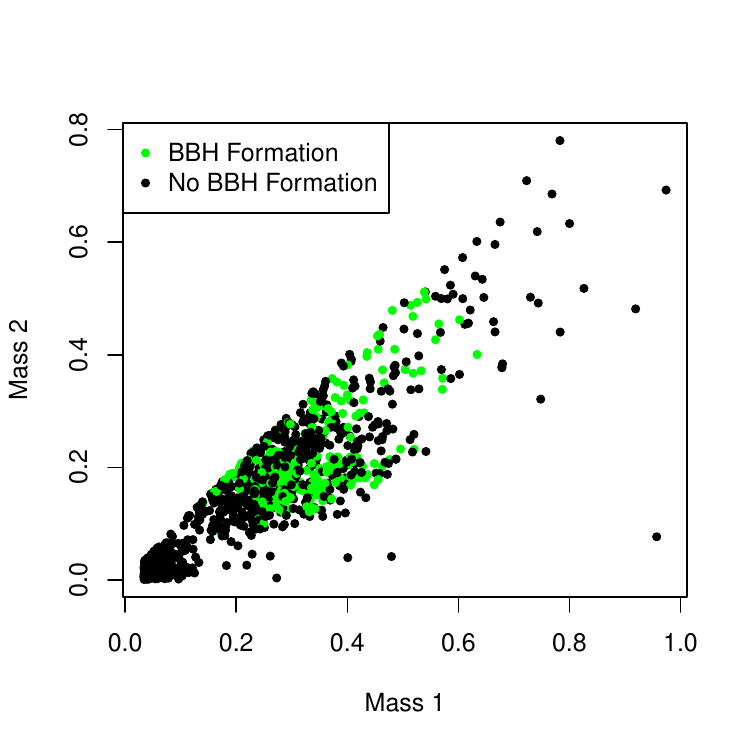}
\hspace{1cm}
\includegraphics[scale=0.5,trim=-10 10 0 40]{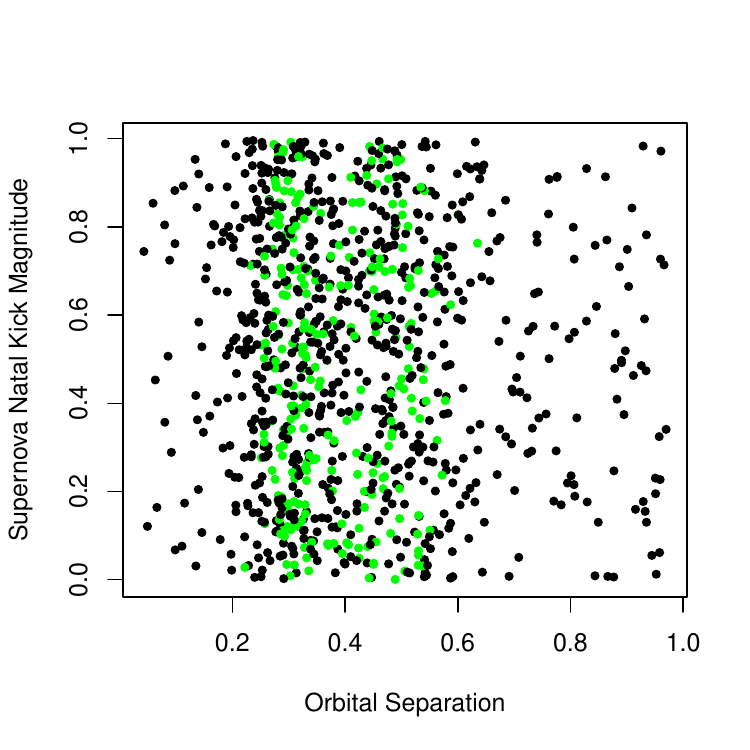} 
\caption{COMPAS simulations projected onto two pairs of inputs. Green dots indicate BBH formations.}
\label{fig:bbh_eda}
\end{figure}
Figure \ref{fig:bbh_eda} illustrates the relationship between some of the
inputs and BBH formation. The left panel projects a subset of evaluations
collected by \cite{Lin2021} onto the two celestial mass inputs, where the
first object denotes the more massive of the two. The right panel projects the
same data onto the orbital separation and supernova natal kick magnitudes.
Observe in both plots how BBH formations occur in relatively small sub-regions
of the input space. For instance, orbital separation values outside of
$[0.2,0.6]$ or greater than $0.6$ almost never yield a BBH.

We use simulations from the COMPAS provided by \cite{Lin2021}, which involved
using adaptive importance sampling to boost the percentage of BBH formations
to around $27\%$. We randomly subsampled these data into ten unique training
sets of size $n=10{,}000$, then fit GP-based classifiers [Section
\ref{sec:imp_details}] for $1{,}000$ testing locations randomly drawn from the
remaining samples.
\begin{figure}[ht!]
\centering
\includegraphics[scale=0.5,trim=0 22 0 0,clip=TRUE]{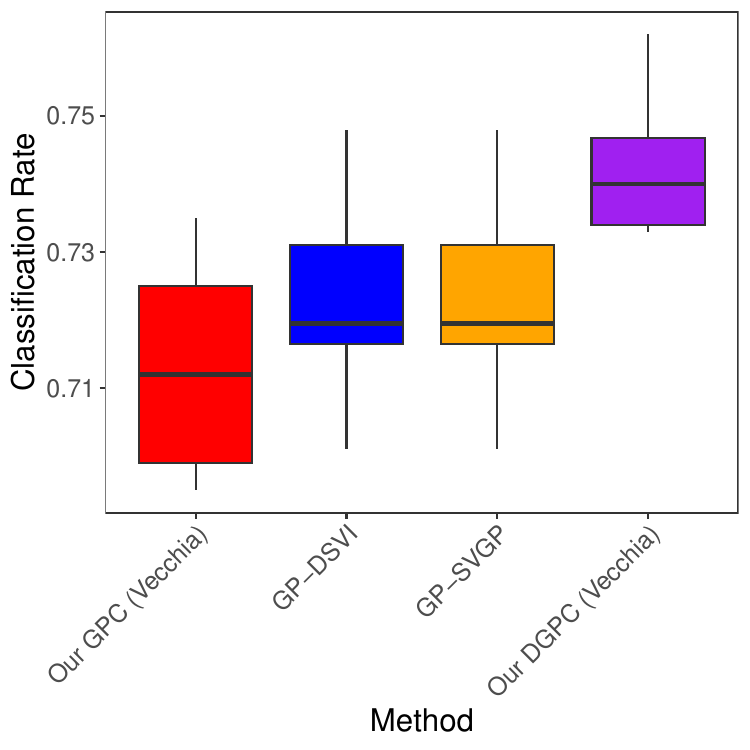}
\hspace{1cm}
\includegraphics[scale=0.5,trim=0 22 0 0,clip=TRUE]{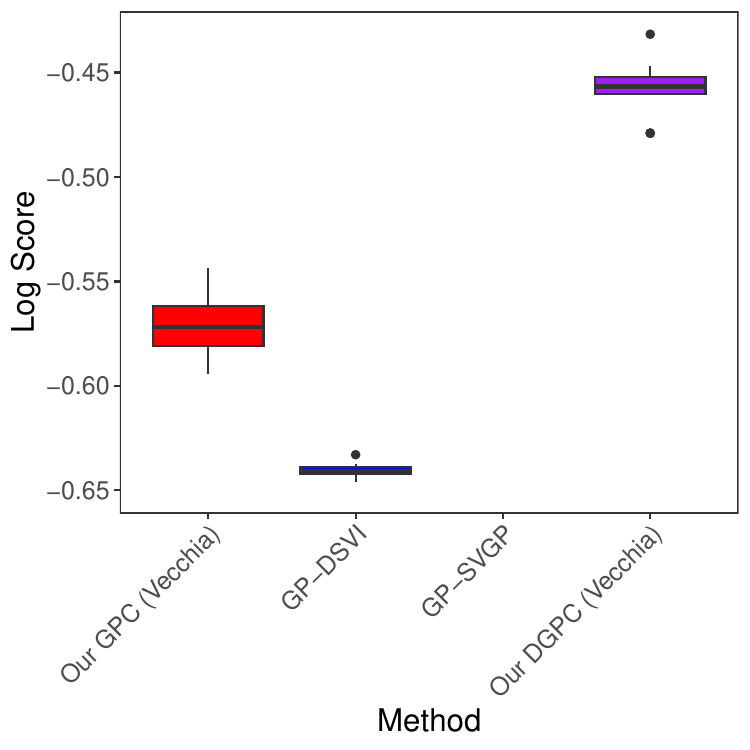}
\caption{BBH comparison. 
SVGP's LSs are out-of-frame, being well below the other methods.}
\label{fig:bbh_class_results}
\end{figure}
Results are shown in Figure \ref{fig:bbh_class_results}. In terms of pure
accuracy (CR), all three models perform similarly. While DSVI and SVGP achieve
slightly better CR, our GPC has the highest median CR and has less variability
across the ten runs. What's more striking is the comparison on LS.  SVGP's are
so poor, they fall below the $y$-axis and are not shown. Our Vecchia GPC
obtained the best LS in all ten of the MC instances.

\section{Deep Gaussian process classification}
\label{sec:deepgp}

Most GPs make an assumption of \textit{stationarity}; i.e., that modeling --
via the MVN covariance structure -- depends only on relative input distance,
not input position.  A nonstationary regression surface (i.e., with
real-valued outputs) is easy to describe as having disparate regimes in the input
space, say one region where outputs are wiggly and another that's changing
more gradually. Nonstationarity in classification is more subtle. The
deterministic nature of simulators like the BBH, or any of the synthetic
examples we have explored so far, means transitions between output classes 
are always abrupt.

\begin{figure}[ht!]
\centering
\includegraphics[scale = 0.5,trim=15 10 20 25]{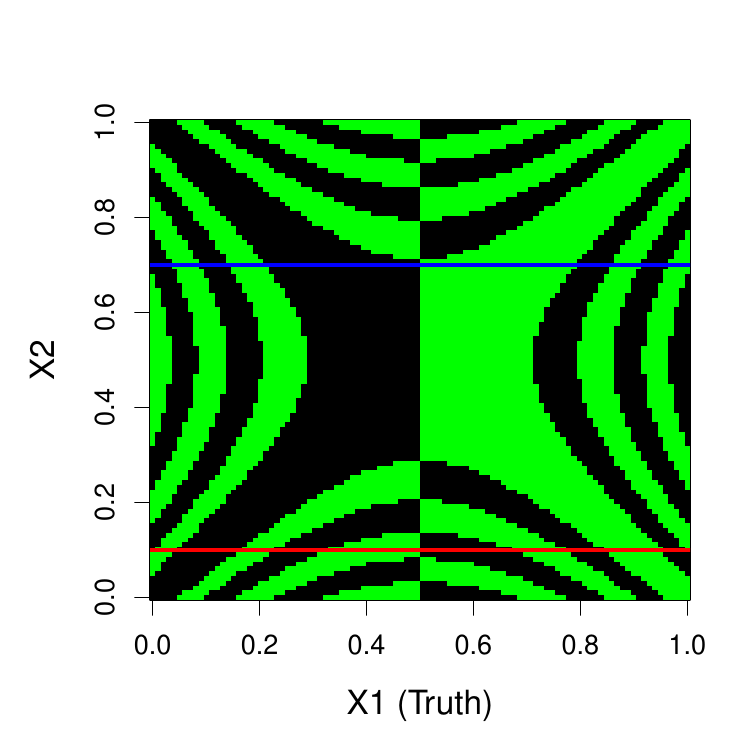}
\includegraphics[scale = 0.5,trim=5 10 20 25]{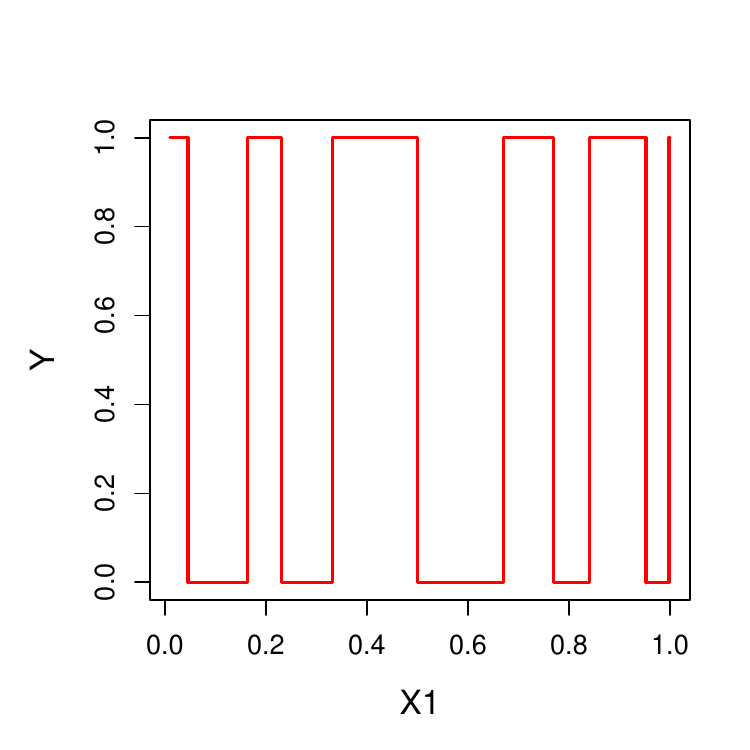}
\includegraphics[scale = 0.5,trim=5 10 30 25]{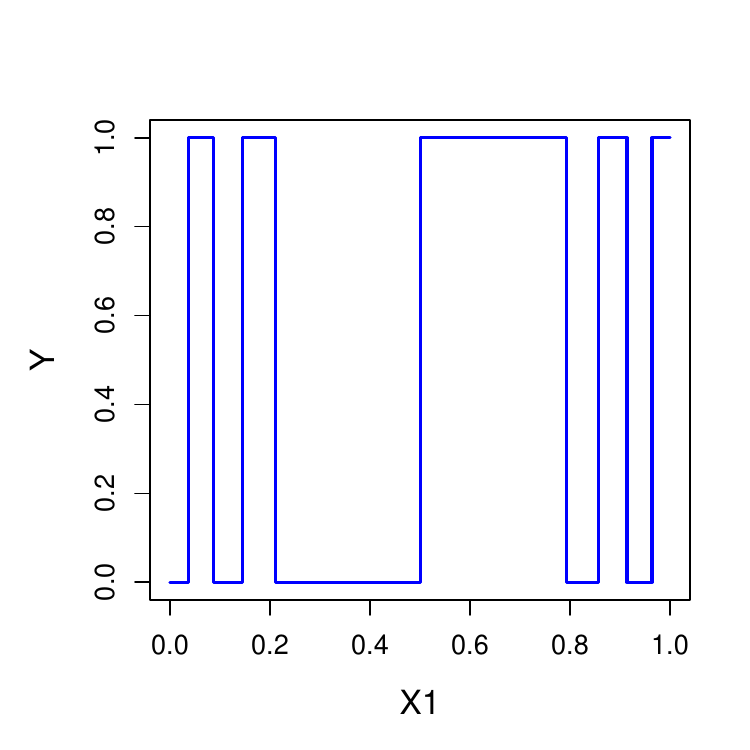}
\caption{2d Schaffer function ({\em left}) with slices along
$x_2=0.1$ ({\em middle}, red) and $x_2 = 0.7$ ({\em right}, blue).}
\label{fig:nonstationary}	  
\end{figure}
To illustrate the nuances of nonstationarity in a classification setting, 
consider the Schaffer function from Section \ref{sec:sim_ex},
re-depicted in the left panel of Figure \ref{fig:nonstationary}.
Two horizontal slices are shown in the middle and right panels,
in red and blue, respectively. Notice how
transitions are more evenly spaced in the first slice than the second.
Broadly, labels change less rapidly in the middle of the space than 
near the edges/corners.  Since the behavior of the response is affected
by both position and relative distance, our GP classifier would benefit
from additional nonstationary flexibility.

\subsection{DGPC methodology and implementation}

For real-valued, regression GPs, there are many approaches to relaxing
stationarity, including process convolutions \citep{higdon1998process} and
spectral kernels \citep{remes2017non}.  For a comprehensive overview, see
\citet{sauer2023non}. One increasingly popular strategy was formerly known as
{\em input warping} \citep{sampson1992nonparametric, schmidt2003bayesian}.
The idea is to adjust the inputs so that stationary modeling is appropriate.
\citeauthor{sampson1992nonparametric} saw this as a pre-processing step,
whereas \citeauthor{schmidt2003bayesian} considered joint inference for {\em
latent}, warped inputs $(X \rightarrow W)$ and downstream input--output $(W
\rightarrow Z)$ modeling.  Placing a GP prior on both mappings creates a
functional composition much like a two-layer neural network (NNs), but with
GPs.  Piggybacking off of buzz for deep NNs, \citet{Damianou2012} rebranded
the idea as deep GPs (DGPs), entertaining further (deeper) layers of
composition.

\citeauthor{Damianou2012}'s main contribution involved modernizing inference
for the latent, warped ``layers'', $W$.  \citeauthor{schmidt2003bayesian}'s
Metropolis approach suffers the same limitations outlined earlier [Section
\ref{sec:class_review}] for latent $Z$-variables in a classification context,
limiting applicability to small data sizes. \citeauthor{Damianou2012}'s VI/IP
allowed for much larger $n$, in neat symmetry to the situation we described
earlier for classification with GPs: cumbersome sampling replaced by
optimization and approximation. But like in that setting, computational gains
come at the expense of fidelity and UQ.

\citet{Sauer2020} recognized that ESS is ideal for fully Bayesian, posterior
integration of latent warping variables.  With the right inferential mechanism
(ESS), and covariance approximation (Vecchia), two-layer DGPs were sufficient
for modeling nonstationary simulation dynamics \citep{Sauer2022}.  We
appropriated many of those ideas for classification in Section
\ref{sec:method} in a one-layer, shallow GPC context.  Here we chain the two
together, bolting an additional latent Bernoulli layer onto a DGP regression
for classification (DGPC).  This is easy to say, but hard to do, which may be
why it has yet to be done.  

Let $W\in\mathbb{R}^{n\times d}$ represent a $d$-dimensional warping layer
situated between $X \in \mathbb{R}^{n\times d}$ and logistic latent $Z \in
\mathbb{R}^n$ variables. While $W$ can be of any column dimension, it is perhaps most
intuitive to match with $X$.  \citet{Sauer2020} showed that using a lower
dimensional warping, providing an {\em autoencoder}-like effect by analogy to
DNNs \citep{kingma2022autoencodingvariationalbayes}, may have deleterious
performance for computer experiments.  
\begin{figure}[ht!]
\begin{align*}
\vcenter{\hbox{\includegraphics[scale=0.35]{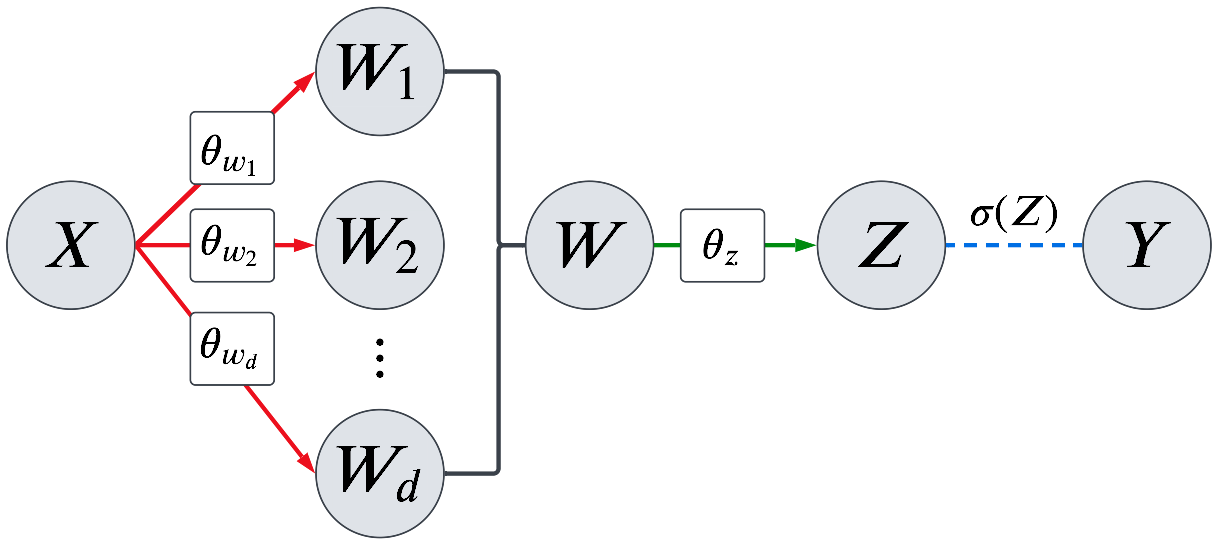}}}
\qquad\qquad
\begin{aligned}
\mathcolor{blue}{Y\mid Z &\sim \text{Bern}\left(\sigma(Z)\right)}\\
\mathcolor{Green}{Z\mid W &\sim \mathcal{N}_n\left(0, \Sigma_{\theta_z}(W)\right)}\\
\mathcolor{red}{W_j\mid X &\stackrel{\mathrm{ind}}{\sim} \mathcal{N}_n\left(0, 
    \Sigma_{\theta_{w_j}}(X)\right) \quad j=1,\dots,d}
\end{aligned}
\end{align*}
\caption{{\em Left:} Diagram of two-layer
DGPC. Each arrow corresponds to a separate GP. {\em Right:} DGPC model
hierarchy, color-coded to match components of the diagram.}
\label{fig:dgp_diagram}
\end{figure}
The architecture of our model is diagrammed and detailed hierarchically in
Figure \ref{fig:dgp_diagram}. Observe the solid red arrows from $X$ to the
coordinates of $W$; each represents a separate, independent GP (see the
equation in red to the right). These collectively map $X$ to a warped version
of itself ($W = [W_1, W_2, \dots, W_d]$), which 
is fed as input to the latent GP for $Z$ (green arrow), then
logistically-transformed to a success probability for $Y$ (blue dotted line).
We introduce subscripts $\theta_z$ and $(\theta_{w_1},\dots,\theta_{w_d})$ 
to denote the covariance matrices constructed with independent lengthscales 
for $Z$ and each coordinate of $W$, respectively.

Algorithm \ref{alg:gibbs2} outlines our Gibbs procedure for DGPC. Samples for
each $\theta_{w_j}$ are generated through MH via the MVN likelihood
$\mathcal{L}\left(W_j|X,\theta_{w_j}\right)$. Posterior sampling for each
coordinate of $W$ follows ESS \citep{Sauer2020}. Specifically, take
$W_j\sim\mathcal{N}_n\left(0, \Sigma_{\theta_{w_j}}(X)\right)$, then convolve
with $W_j^{(t)}$ via a random angle $\gamma$ to propose $W_j^\star$
[Alg.~\ref{alg:ess}]. Acceptance is based on the likelihood ratio
$\mathcal{L}(Z\mid W^\star,\theta_z)/\mathcal{L}(Z\mid W^{(t)},\theta_z)$,
with only the $j^\text{th}$ column of $W^\star$ and $W^{(t)}$ differing at a
time. Finally, posterior samples of $Z$ can be obtained by substituting
$W^{(t)}$ in place of $X$ in the ESS procedure outlined in Alg.~\ref{alg:ess}.
Altogether, Alg.~\ref{alg:gibbs2} is similar to Alg.~\ref{alg:gibbs_binary}
with additional loops for $W_j$ and its lengthscale(s).

\medskip
\begin{algorithm}[H]        
\DontPrintSemicolon
Initialize $\theta_{w_1}^{(1)},\dots,\theta_{w_d}^{(1)}$, 
$\theta_{z}^{(1)}$, 
$W^{(1)}$, $Z^{(1)}$\;
\For{$t = 2, \dots, T$}{
    \For{$j = 1, \dots, d$}{
    $\theta_{w_j}^{(t)} \sim \pi(\theta_{w_j}\mid X,  W_j^{(t-1)})$ 
    \tcp*{MH via $\mathcal{L}(W_j\mid X, \theta_{w_j})$}
    $W_j^{(t)} \sim \pi(W_j \mid X, Z^{(t-1)}, \theta_{w_j}^{(t)})$ 
        \tcp*{ESS via $\mathcal{L}(W_j \mid X, \theta_{w_j}), 
            \mathcal{L}(Z^{(t-1)} \mid W, \theta_{z})$}
    }
    $\theta_z^{(t)} \sim \pi(\theta_z\mid W^{(t)}, Z^{(t-1)})$ 
    \tcp*{MH via $\mathcal{L}(Z\mid W, \theta_z)$; like Alg.~\ref{alg:gibbs_binary} with $W$ for $X$}
    $Z^{(t)} \sim \pi(Z \mid W^{(t)}, Y, \theta_z^{(t)})$ 
        \tcp*{ESS via $\mathcal{L}(Z\mid W,\theta_z), \mathcal{L}(Y \mid Z)$; like Alg.~\ref{alg:gibbs_binary} with $W$ for $X$}}
\caption{Gibbs sampling procedure for two-layer DGPC estimation.}
\label{alg:gibbs2}
\end{algorithm}
\medskip

Predictive quantities for new inputs $\mathcal{X}$ involve additional MC
sampling like with the ordinary GPC. Posterior draws for $\mathcal{Z}$ require
feeding samples from Alg.~\ref{alg:gibbs2} through a cascade of kriging
equations (\ref{eq:pred}).  For example, take
$\mathcal{W}_j^{(t)}(\mathcal{X})\mid X,W_j^{(t)}$ by replacing $Y$ with
$W_j^{(t)}$ and using $\theta_{w_j}^{(t)}$ within each $\Sigma(\cdot)$.
Combine these samples for all $j$ to form $\mathcal{W}^{(t)} =
\left[\mathcal{W}_1^{(t)},\dots,\mathcal{W}_d^{(t)}\right]$.
Then form $\mathcal{Z}^{(t)}\left(\mathcal{W}^{(t)}\right)\mid W^{(t)},
Z^{(t)}$ in a similar fashion. From there we can approximate the mean and
variance of $p_y(x)$ for each $x\in\mathcal{X}$ just as for ordinary GPC
(\ref{eq:post_pred}).  Introducing $d$ additional GPs compounds the
computational burden for training and prediction, especially when $d$ is large
like for the BBH. Yet the same Vecchia approximations are applicable, both for
inference and prediction.  Implementation is similar to that of Section
\ref{sec:implement}, via separate randomized orderings for each of the $d+1$
GPs.  The entire procedure remains cubic in the conditioning set size, $m$.

\subsection{DGPC performance comparison}

First consider a two-layer deep GPC on the simulated examples from Section
\ref{sec:sim_ex}.
\begin{figure}[ht!]
\centering
\begin{minipage}{7.5cm}
\includegraphics[scale=0.6,trim=-10 10 0 40]{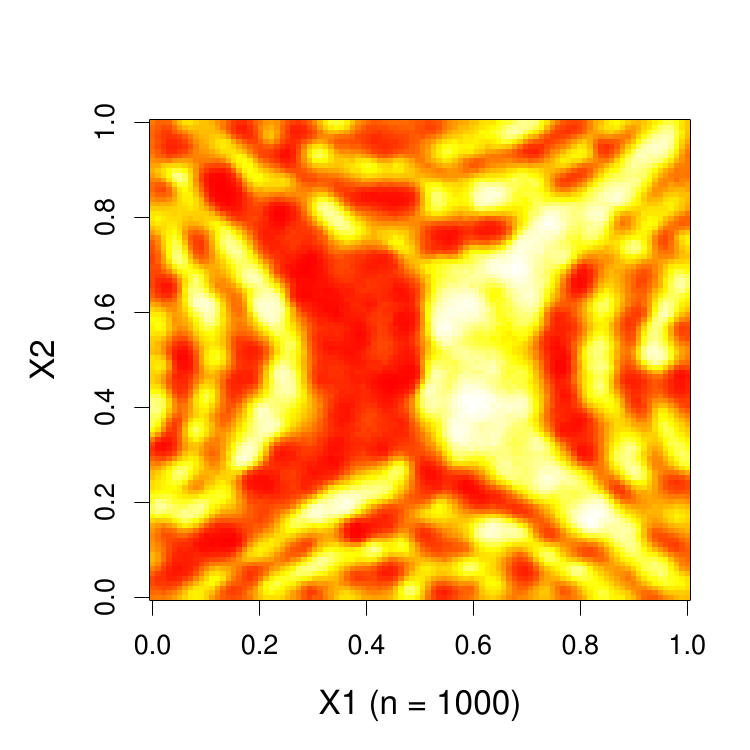}
\end{minipage}
\begin{minipage}{7.5cm}
\includegraphics[scale=0.6,trim=-10 10 0 40]{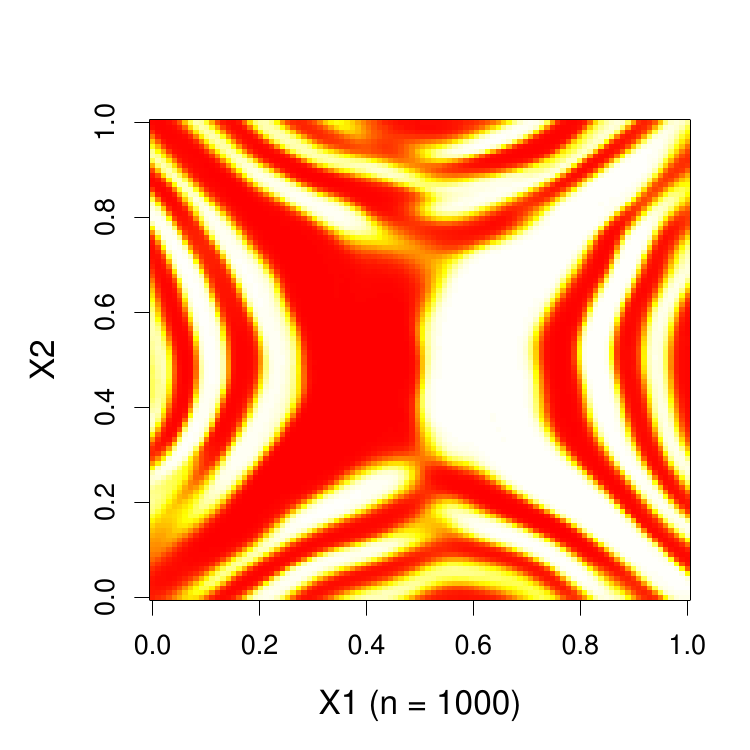} 
\end{minipage}
\caption{Vecchia GPC ({\em left}) and DGPC ({\em right}) on  
2d Schaffer no.~4 with the identical training points.}
\label{fig:dgp_comp}
\end{figure}
Figure \ref{fig:dgp_comp} demonstrates improved predictive resolution for the
2d Schaffer no.~4 example. The left panel shows the posterior mean success
probability surface from our ordinary GPC trained on an LHS design of size
$n=1{,}000$; the right panel shows a two-layer DGPC on the same data. Notice
how the DGPC provides sharper transitions between labels, especially at the
edges/corners. Moreover, it furnishes more confident probability estimates in
regions of a single class, as seen by the brighter colors in the middle.

Out-of sample performance for DGPC is provided by the purple boxplots contained
in Figures \ref{fig:schaffer_results} and \ref{fig:g6_results}, for the
Schaffer and ``G'' functions, respectively. In both cases, the additional
depth of the DGPC leads to a significant improvement in accuracy (CR) and UQ
(LS), especially for smaller $n$. With larger $n$, additional depth provides
diminishing returns. Ultimately, both GP and DGP are universal approximators.
DGPC performance on BBH classification is provided by the purple boxplots in
Figure \ref{fig:bbh_class_results}.  It provides both the highest median
classification accuracy and the most effective UQ (as indicated by high LS).

\section{Discussion}
\label{sec:discuss}

We have deployed the Vecchia approximation to make fully Bayesian GP
classification (GPC) feasible in large-data settings.  By speeding up matrix
decomposition through sparsity-inducing approximations, which fit naturally
into the previously intractable elliptical slice sampling framework for latent
GP classification, we have allowed for GPC training in large data settings
without sacrificing full uncertainty quantification (UQ). This has led to
improved accuracy and UQ, as demonstrated on a suite of benchmark problems
and a real binary black hole simulator. Additionally, we have shown how
additional input-warping layers could be accommodated similarly, resulting in
a deep GP classifier, for nonstationary classification.

While our work in this paper focuses on binary classification, the extension
to more than two classes is straightforward \citep{Rasmussen2006}. Inference
for a response with $K$ classes can be performed by expanding $Z$ to $K-1$
independent GP layers \citep{seeger2004}. Let $Z_k\sim\mathcal{N}\left(0,
\Sigma_k\left(X\right)\right)$ for $k=1,\dots,K-1$, where each $\Sigma_k(X)$
contains its own covariance hyperparameters. The \textit{generalized} logistic
function transforms this layer into valid probabilities: $\sigma(z_{ik}) =
\dfrac{e^{-z_{ik}}}{1 + \sum_{\ell=1}^{K-1} e^{-z_{i\ell}}}$, where $z_{ik}$
represents the $i^{\mathrm{th}}$ entry of the $k^{\mathrm{th}}$ layer of $Z$.
When $K=2$ this function simplifies to the sigmoid inverse-link function seen
in Section \ref{sec:class_review}. Our publicly available implementation in
\textsf{R}\footnote{\url{bitbucket.org/gramacylab/deepgp_class}} supports
multiclass classification.

Previous research on COMPAS/BBH has focused on regression for predicting
chirp mass. \cite{Lin2021} propose a local classification GP to predict BBH
formation where inputs predicted as non-{\tt NA} feed into a chirp mass
surrogate. However, their point-estimation approach may undercut on UQ. 
Combining our approach with a fully Bayesian regression GP may be an interesting
avenue for further research.  An alternative model, recently proposed by
\cite{yazdi2024deep}, skips the classification step to directly model {\tt NA}
values as zero with a DGP.  It would be of interest to contrast these two
approaches.
	
\bibliography{class}

\newpage
\appendix

\section{2d box results}
\label{appendix:box}

\begin{figure}[ht!]
\centering
\includegraphics[scale=0.5,trim=20 0 0 0]{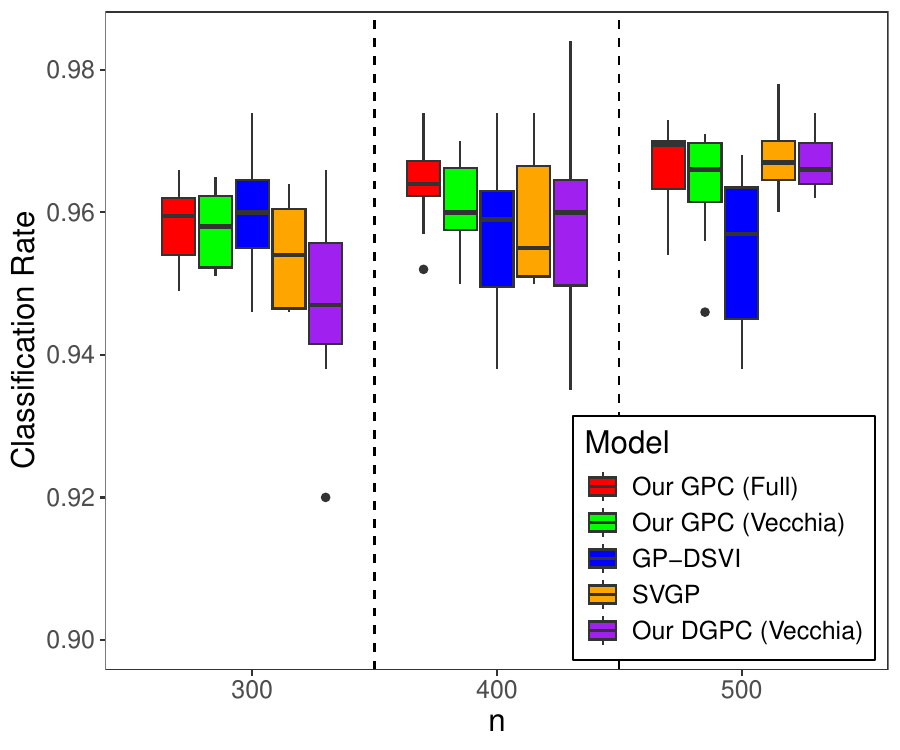}
\hspace{1cm}
\includegraphics[scale=0.5,trim=0 0 0 0]{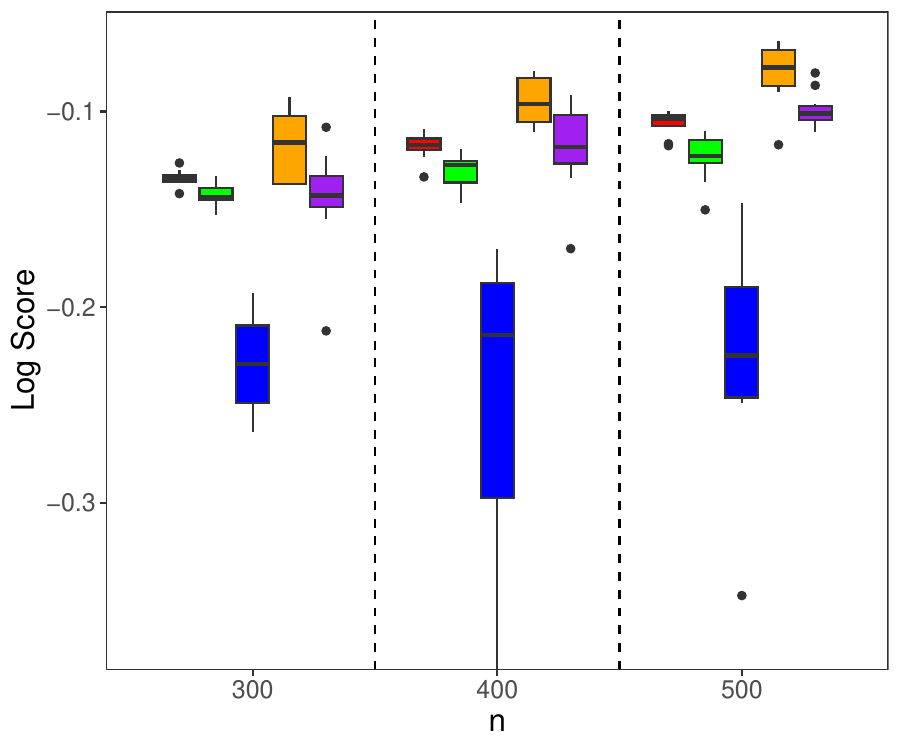}
\caption{Two-dimensional ``box'' results. Some negative outliers for DSVI's log score are cut off from the plot to make the rest of the boxplots more readable.}
\label{fig:box_results}
\end{figure}
Results for the two-dimensional box example are presented via boxplots in Figure \ref{fig:box_results}. The
full GPC slightly outperforms the Vecchia-approximated GPC (albeit
at higher computational expense). DSVI performs poorly across the board.
SVGP performs slightly better on average for training sizes of $n=500$, but has
slightly larger variance than our method in both classification rate and
log score. Vecchia-DGPC only outperms our ordinary GPC when $n=500$, indicating more data is needed for the benefits of adding deepness to be readily apparent. Note, a few lower outliers from DSVI in both classification rate and
log score are not shown for plot scaling purposes to highlight the performance
of the rest of the methods.

\end{document}